\documentclass[12pt]{iopart}

\pdfoutput=1

\usepackage[english]{babel}
\usepackage{
amssymb,amsfonts,latexsym}
\usepackage{graphicx}
\usepackage{color}
\usepackage{iopams}
\usepackage{subfigure}
\usepackage{afterpage}
\usepackage{blkarray}
\usepackage{cite}
\usepackage{tikz}
\usetikzlibrary{snakes}
\usetikzlibrary{calc}
\usepackage{bm}

\definecolor{blue(pigment)}{rgb}{0.2, 0.2, 0.6}
\usepackage{mathrsfs}
\usepackage{times}
\usepackage[colorlinks,
citecolor=blue,linkcolor=blue,urlcolor=blue]{hyperref}
\usepackage{scalerel}
\usepackage{tensor}
\usepackage{etoolbox}

\makeatletter
\def\@mkboth#1#2{}
\newlength\appendixwidth
\preto\appendix{\addtocontents{toc}{\protect\patchl@section}}
\newcommand{\patchl@section}{%
  \settowidth{\appendixwidth}{\textbf{Appendix }}%
  \addtolength{\appendixwidth}{1.5em}%
  \patchcmd{\l@section}{1.5em}{\appendixwidth}{}{\ddt}%
}
\makeatother

\def\eqref#1{(\ref{#1})}

\usepackage{cancel}

\newcommand{\be}{\begin{equation}}
\newcommand{\ee}{\end{equation}}
\newcommand{\bea}{\begin{eqnarray}}
\newcommand{\eea}{\end{eqnarray}}

\newcommand{\fa}{\mathfrak{a}}
\newcommand{\fb}{\mathfrak{b}}

\newcommand\reallywidehat[1]{\arraycolsep=0pt\relax%
\begin{array}{c}
\stretchto{
  \scaleto{
    \scalerel*[\widthof{\ensuremath{#1}}]{\kern-.5pt\bigwedge\kern-.5pt}
    {\rule[-\textheight/2]{1ex}{\textheight}} 
  }{\textheight} %
}{0.5ex}\\           
#1\\                 
\rule{-1ex}{0ex}
\end{array}
}

\begin{document}

\title[Loschmidt echo in $XXZ$ Heisenberg spin chains]{From the
  Quantum Transfer Matrix to the Quench Action: The Loschmidt echo in $XXZ$ Heisenberg spin chains }
\author{Lorenzo Piroli$^{1}$, Bal\'{a}zs Pozsgay$^{2,3}$, Eric Vernier$^{1}$}
\address{$ˆ1$ SISSA and INFN, via Bonomea 265, 34136 Trieste, Italy}
\address{$^2$ Department of Theoretical Physics, Budapest University
  of Technology and Economics, 1111 Budapest, Budafoki \'{u}t 8, Hungary}
\address{$ˆ3$ MTA-BME ``Momentum'' Statistical Field Theory Research Group, 1111 Budapest, Budafoki \'{u}t 8, Hungary}

\ead{lpiroli@sissa.it, pozsgay.balazs@gmail.com, evernier@sissa.it}
\date{\today}

\begin{abstract}
We consider the computation of the Loschmidt echo after quantum
quenches in the interacting $XXZ$ Heisenberg spin chain both for real
and imaginary times. We study two-site product initial states,
focusing in particular on the N\'eel and tilted N\'eel states.
We apply the Quantum Transfer Matrix (QTM) approach to derive generalized TBA equations, which follow
from the fusion hierarchy of the appropriate QTM's. Our formulas are valid for arbitrary imaginary time and for real times at least up to a time $t_0$, after which the integral equations have
to be modified. In some regimes, $t_0$ is seen to be either very large
or infinite, allowing to explore in detail the post-quench dynamics of
the system. As an important part of our work, we show that for the N\'eel state our imaginary time
results can be recovered by means of the quench action approach,
unveiling a direct connection with the quantum transfer matrix
formalism. In particular, we show that in the zero-time limit, the study of our TBA equations allows for a simple alternative derivation of the
recently obtained Bethe ansatz distribution functions for the N\'eel,
tilted N\'eel and tilted ferromagnet states.

\end{abstract}


\maketitle

{\hypersetup{linkcolor=blue(pigment)}
\tableofcontents
}

\section{Introduction}

One of the main motivations in the study of integrable systems is the possibility to provide reliable analytic results in the investigation of physically relevant but theoretically challenging problems. As a consequence, these models have traditionally played the role of the perfect benchmark to test the range of validity of approximate or perturbative methods developed to study more complicated systems.

This point of view has rapidly evolved in the past decade, due to the revolutionary progress in cold atomic physics \cite{bdz-08,ccgo-11,pssv-11,gbl-13}. Indeed, a renewed motivation in the study of integrable models has come by their direct relevance for cold atomic experiments, where simple fine-tuned Hamiltonians can be realized with high control and physical quantities measured to high precision. Of particular importance is the possibility to experimentally realize different non-equilibrium settings \cite{gmhb-02,kww-06,cetal-12,gklk-12,shr-12,fse-13,lgkr-13,mmkl-14,glms-14}; among these the simplest is arguably that of a quantum quench, where a system is prepared in a well defined initial state and let evolved unitarily according to a known Hamiltonian \cite{cc-06}.

These developments have unraveled the need for a solid theoretical understanding of key aspects of non-equilibrium dynamics in isolated quantum systems, and an active field of research has focused on the physical consequences of integrability. Of particular interest has been the question of whether and how a system locally approaches a stationary state at long times after the quench. Among the many results, excellent studies have now established the validity of a generalized Gibbs ensemble to predict the steady asymptotic value of local correlations \cite{rdyo-07,cazalilla-06,rigol-09,cdeo-08,fm-10,
cef-11,fe2-13,rf-11,ck-12,sc-14,pozsgay2-13,fe-13,fcec-14,
wdbf-14,wdbf-14_2,pmwk-14,pozsgay2-14,ga-14,idwc-15,iqdb-16,pvc-16,gkf-16,iqc-16}, as well as the existence and physical importance of quasi-local conserved charges in integrable systems \cite{prosen-11,ip-12,prosen-14,zmp-16, imp-15,ppsa-14,pv-16,fagotti-14,
fagotti-16,emp-15,dlsb-15,bs-16,vecu-16,cardy-16,doyon-15}. For a pedagogical introduction to these topics see the recent reviews \cite{ef-16,impz-16,vr-16,dm-16,cc-16}.

It is hard to overemphasize the importance of analytical methods in these theoretical achievements. A significant role has been played by the introduction of the so called quench action approach, which has proven to be a powerful tool in the study of quantum quenches in Bethe ansatz integrable models \cite{ce-13,caux-16}. Indeed, the latter has already led to beautiful results in the characterization of post-quench steady states in several classes of experimentally relevant non-equilibrium settings \cite{dwbc-14,wdbf-14,wdbf-14_2,pmwk-14,bse-14,dmv-15,pce-16,bpc-16,bucciantini-16,alca-16}.

Despite the many successes, the computation of the whole time evolution following a quantum quench still represents a remarkable challenge. While by now several non-trivial explicit calculations have been performed in systems which can be mapped onto free ones \cite{cazalilla-06,bpgd-09,cef-11,mc-12,se-12,csc-13,kcc-14,rs-14,bkc-14,dc-14,msca-16,bf-16}, only a few analytical or semi-analytical results are available for quenches in genuinely interacting models \cite{ia-12,sg-72, bse-14,mussardo-13,delfino-14,dpc-15,pe-16,cubero-16,vwed-16,kz-16,ac-16,zl-16}. The development of analytical techniques of investigations complementary or beyond the present methods thus represents an urgent issue.

Among the physically interesting quantities to probe the post-quench dynamics, the simplest is arguably the Loschmidt echo, namely the squared absolute value of the overlap between evolved and initial states. The latter has recently received a lot of attention, especially for its relevance in studies of dynamical phase transitions \cite{qslz-06,silva-08,pmgm-10,sd-11,fagotti2-13,hpk-13,dpfz-13,ks-13,pozsgay-13,
cwe-14,heyl-14,vths-13,as-14,cardy-14,deluca-14,ps-14,kk-14,vd-14,
heyl-15,ssd-15,sdpd-16,zhks-16,ps-16,zy-16}. Despite its simplicity, even for this quantity analytical results in interacting models are still extremely hard to obtain when a mapping onto free systems is not possible.

An analytical computation of the "Loschmidt echo per site" in the interacting $XXZ$ Heisenberg spin chain was presented in \cite{pozsgay-13}, where quantum quenches from the initial N\'eel state were studied by means of a quantum transfer matrix approach. Originally developed in the investigation of finite temperature problems \cite{klum-93,suzuki-99, klumper-04}, its extension to the computation of the Loschmidt amplitude relies on the interpretation of the latter as a particular boundary partition function. This natural identification has been exploited many times in the literature \cite{silva-08,hpk-13,ks-13,cwe-14,vths-13,as-14,cardy-14,ps-14,vd-14,ssd-15,sdpd-16}. In the particular case of $XXZ$ Heisenberg chains, this idea was also at the basis of the recent work \cite{as-14}, where an efficient numerical approach allowed to perform a remarkably detailed analysis of the Loschmidt echo in the thermodynamic limit for many different quantum quenches.

In this paper we build upon the work \cite{pozsgay-13} and provide further results for a fully analytical computation of the Loschmidt echo in the XXZ Heisenberg chain. We first revisit the case of the initial N\'eel state considered in \cite{pozsgay-13} and provide alternative formulas for the Loschmidt echo per site both at real and imaginary times (which is directly related to the so called dynamical free energy \cite{pozsgay-13, fagotti2-13}). These involve the solution of integral equations obtained by the so-called fusion relations of boundary transfer matrices. 
The formulas presented in this work allow for an explicit efficient numerical evaluation of the Loschmidt echo for real times.

Going further, we show that the same approach can be applied straightforwardly in the case of more general two-site product states such as tilted ferromagnets and tilted N\'eel states. We provide in particular a detailed study of the latter case, for which a corresponding set of integral equations is derived. We argue that these are valid for arbitrary imaginary times, while for real times only up to a time $t_0$, which depends on the initial state and the final Hamiltonian parameters. In some regimes $t_0$ is seen to be either very large or infinite, allowing to probe in detail the relaxation dynamics of the system. For times larger than $t_0$ our formulas have to be modified in a way which is in principle feasible and discussed in the following sections. 

Finally, we consider the computation of the Loschmidt echo per site at imaginary times by means of the quench action approach. In the case of the N\'eel state we show that the quantum transfer matrix results can be recovered, unveiling a direct link between the two approaches. Based on this, we provide an alternative derivation of the recently obtained Bethe ansatz rapidity distribution functions corresponding to the tilted ferromagnet and tilted N\'eel states \cite{pvc-16}. Our treatment also naturally clarifies the validity of certain analytical properties of the latter (namely, the so called $Y$-system relations), which were previously established numerically \cite{iqdb-16, pvc-16}.

The rest of this article is organized as follows. In section~\ref{sec:setup} we introduce the $XXZ$ Heisenberg chain and the quench protocol. In section~\ref{sec:qtm} we review the quantum transfer matrix construction, while in section~\ref{sec:boundary_bethe} some aspects of the boundary algebraic Bethe ansatz are presented. The analytical results for the Loschmidt echo at imaginary times are derived in section~\ref{sec:TBA_equations}, and continuation to real times is discussed in section~\ref{sec:resolution}, where our analytical formulas are explicitly evaluated numerically. Section~\ref{sec:qam} is devoted to the comparison of the quantum transfer matrix formalism with the quench action approach. Finally, we report our conclusions in section~\ref{sec:conclusion}, while some technical aspects of our work are provided in the appendices.

\section{Setup}\label{sec:setup}

We consider the $XXZ$ spin-$1/2$ Heisenberg model, defined on a chain
of length $L$. Since we will be interested in quenches
where the initial state is a two-site product state, we will always assume $L$ to be an even
number. The Hamiltonian of the model reads 
\ \\
\ \\
\bea
H &=& J \sum_{j=1}^{L}\left[s^{x}_j s^{x}_{j+1}+  s^{y}_js^{y}_{j+1}+ \Delta \left( s^{z}_js^{z}_{j+1}-\frac{1}{4}\right)\right] \,\nonumber\\
&=& \frac{J}{4} \sum_{j=1}^{L}\left[\sigma^{x}_j \sigma^{x}_{j+1}+  \sigma^{y}_j\sigma^{y}_{j+1}+ \Delta \left( \sigma^{z}_j\sigma^{z}_{j+1}-1\right)\right] \,,
\label{hamiltonian}
\eea
where we take $J>0$, while $s_j^{\alpha}=\sigma^{\alpha}_j/2$ are the local spin operators,  $\sigma_j^{\alpha}$ being the Pauli matrices. We further assume periodic boundary conditions, $\sigma_{L+1}^\alpha \equiv \sigma_1^\alpha$, and restrict to the gapped antiferromagnetic phase $\Delta>1$. We then introduce the parametrization  
\be
\Delta= \cosh \eta\,, 
\label{eta_parameter}
\ee
with $\eta \in \mathbb{R}$. The Hamiltonian \eqref{hamiltonian} can be diagonalized by means of the Bethe ansatz method \cite{kbi-93}. Relevant aspects of the latter will be discussed in the next section.

In this work we will consider quantum quenches from two-site product states of the form
\be
|\Psi_0\rangle=|\psi_0\rangle_{1,2}\otimes|\psi_0\rangle_{3,4}\otimes \ldots \otimes |\psi_0\rangle_{L-1,L}=|\psi_0\rangle^{\otimes L/2}\,,
\label{eq:initial}
\ee
where $|\psi_0\rangle \in \mathbb{C}^2\otimes \mathbb{C}^2$ is an arbitrary state. Two relevant examples are the tilted N\'eel state defined by
\be
\fl |\psi_0\rangle=
|\vartheta;\swarrow\nearrow\rangle=
\left[-\sin\left(\frac{\vartheta}{2}\right)|\uparrow\rangle+\cos\left(\frac{\vartheta}{2}\right)|\downarrow\rangle\right]
\otimes
\left[\cos\left(\frac{\vartheta}{2}\right)|\uparrow\rangle+\sin\left(\frac{\vartheta}{2}\right)|\downarrow\rangle\right]\,,
\label{tilted_neel}
\ee
and the tilted ferromagnet given by
\be
\fl |\psi_0\rangle=
|\vartheta;\nearrow\nearrow\rangle=\left[\cos\left(\frac{\vartheta}{2}\right)|\uparrow\rangle+\sin\left(\frac{\vartheta}{2}\right)|\downarrow\rangle\right]\otimes \left[\cos\left(\frac{\vartheta}{2}\right)|\uparrow\rangle+\sin\left(\frac{\vartheta}{2}\right)|\downarrow\rangle\right]\,.
\label{tilted_ferromagnet}
\ee
Note that for $\vartheta=0$ the tilted N\'eel state coincides with the well-known N\'eel state 
\be
|N\rangle=|\downarrow\uparrow\ldots\downarrow\uparrow\rangle\,.
\label{eq:neel}
\ee

Given the initial state $|\Psi_0\rangle$, the Loschmidt echo after the quench is defined as
\be
\mathscr{L}(t) =  \left| \langle \Psi_0 | e^{- i  H t} |  \Psi_0 \rangle  \right|^2 \,,
\label{eq:Losch}
\ee
and gives information about the probability of finding the system close to its initial state. For any finite $t$, $\mathscr{L}(t)$ decays exponentially with the volume $L$. It is then useful to define the Loschmidt echo per site \cite{pozsgay-13}
\be
\ell(t) =  \left[\mathscr{L}(t)\right]^{1/L} \,,
\label{eq:losch}
\ee
which is simply related to the so called return rate \cite{as-14}
\be
r(t)=-\frac{1}{L}\log\mathscr{L}(t)=-\log \ell(t)\,.
\label{eq:return}
\ee
As we mentioned in the introduction this quantity has recently received significant attention in the study of dynamical phase transitions, which are associated to points of nonanalyticity of the return rate \eqref{eq:return}. Analytical properties of the latter are more conveniently analyzed by considering a generic complex time and introducing the Loschmidt amplitude
\be
Z(w)=\langle \Psi_0 | e^{-w H } |  \Psi_0 \rangle\,,\qquad w\in\mathbb{C}\,.
\label{eq:partition}
\ee

By interpreting \eqref{eq:partition} as a boundary partition function, it was observed in \cite{hpk-13} that for the transverse Ising chain these nonanalyticities occurred when the system was quenched accross a quantum critical point, establishing a connection between dynamical and equilibrium quantum phase transitions. Subsequent investigations showed that a more complicated picture takes place in general and nonanalyticities can be encountered also for quenches within the same quantum phase \cite{fagotti2-13, vd-14, as-14, ssd-15}.

It turns out that the analytical computation of the Loschmidt amplitude $Z(w)$ is facilitated when $w$ is taken to be a real number, in which case one obtains the so-called dynamical free energy density 
\be
g(w) = \lim_{L\to \infty} \frac{\log Z(w)}{L} \,,\qquad w\in \mathbb{R}\,,
\label{eq:g_en_function}
\ee
which is also connected to the cumulant generating function for the
Hamiltonian \cite{pozsgay-13}. In this work we will consider the
computation of \eqref{eq:g_en_function} for which we provide a full
analytical solution for quantum quenches from initial states of the
form \eqref{eq:initial}. The Loschmidt echo per site (and hence
  the return probablity) is then
given by
\be 
\log \ell(t) = 2 \Re \left[ g(it) \right] \,.
\label{dynfreeenergy}
\ee
We note that \eqref{dynfreeenergy} has to be understood as the
  evaluation of the limit \eqref{eq:g_en_function} for purely
  imaginary $w$ parameters. The $L\to\infty$ limit does not
  necessarily commute with the  $w\to it$ analytic continuation, and
  this is responsible for possible non-analytic behavior of the
  function $g(w)$ in the complex plane. 
As we will see, these issues require a delicate analysis which will be also addressed in our work.

\section{Quantum Transfer Matrix approach to the Loschmidt echo}\label{sec:qtm}

\begin{figure}
\centering
\begin{tikzpicture}
\draw[thick, blue!30 , dashed] (0,0.3) arc (0:180:5 and 0.75);
\draw[gray!50, dashed] (0,0) arc (0:180:5 and 0.75);
\draw[ gray!50, dashed] (0,-0.3) arc (0:180:5 and 0.75);
\draw[ gray!50, dashed] (0,-0.6) arc (0:180:5 and 0.75);
\draw[ gray!50, dashed] (0,-0.9) arc (0:180:5 and 0.75);
\draw[gray!50, dashed] (0,-1.2) arc (0:180:5 and 0.75);
\draw[ gray!50, dashed] (0,-1.5) arc (0:180:5 and 0.75);
 \draw[thick, blue!30 , dashed] (0,-1.8) arc (0:180:5 and 0.75);

   \draw[blue, thick] (0,0.3) arc (0:-180:5 and 0.75); 
\draw (0,0) arc (0:-180:5 and 0.75);\draw[->,>=latex] (0,0) arc (0:-20:5 and 0.75);  
\draw (0,-0.3) arc (0:-180:5 and 0.75);\draw[->,>=latex] (0,-0.3) arc (0:-20:5 and 0.75);  
 \draw (0,-0.6) arc (0:-180:5 and 0.75);\draw[->,>=latex] (0,-0.6) arc (0:-20:5 and 0.75);  
  \draw (0,-0.9) arc (0:-180:5 and 0.75);\draw[->,>=latex] (0,-0.9) arc (0:-20:5 and 0.75);  
   \draw (0,-1.2) arc (0:-180:5 and 0.75);\draw[->,>=latex] (0,-1.2) arc (0:-20:5 and 0.75);  
       \draw (0,-1.5) arc (0:-180:5 and 0.75);\draw[->,>=latex] (0,-1.5) arc (0:-20:5 and 0.75); 
   \draw[blue, thick] (0,-1.8) arc (0:-180:5 and 0.75);

\foreach \angle in {-155,-145,...,-25}
{ 
\draw[->,>=latex] ($(-5,-1.8)+(\angle:5 and 0.75)$) -- ($(-5,0.3)+(\angle:5 and 0.75)$);
}
\node[blue] at   ($(-5,0.55)+(-90:5 and 0.75)$) {$\langle \Psi_0 |$};
\node[blue] at   ($(-5,-2.15)+(-90:5 and 0.75)$) {$| \Psi_0 \rangle$};

\node at (0.75,-0.) {\scriptsize $\beta/2N-\eta$};
\node at (0.5,-0.3) {\scriptsize $-\beta/2N$};
\node at (0.75,-0.6) {\scriptsize $\beta/2N-\eta$};
\node at (0.5,-0.9) {\scriptsize $-\beta/2N$};
\node at (0.75,-1.2) {\scriptsize $\beta/2N-\eta$};
\node at (0.5,-1.5) {\scriptsize $-\beta/2N$};

\node at  ($(-5,-2.15)+(-155:5 and 0.75)$) {\small 1};
\node at  ($(-5,-2.15)+(-145:5 and 0.75)$) {\small 2};
\node at  ($(-5,-2.15)+(-135:5 and 0.75)$) {\small 3};
\node at  ($(-5,-2.15)+(-125:5 and 0.75)$) {\small 4};
\node at  ($(-5,-2.15)+(-115:5 and 0.75)$) {\small \ldots};
\node at  ($(-5,-2.15)+(-55:5 and 0.75)$) {\small \ldots};
\node at  ($(-5,-2.15)+(-35:5 and 0.75)$) {\small L-1};
\node at  ($(-5,-2.15)+(-25:5 and 0.75)$) {\small  L};
\end{tikzpicture}
\caption{Pictorial representation of the quantity $ \langle \Psi_0 | \left[ t(-\beta/2N) t(-\eta + \beta/2N ) \right]^N  |  \Psi_0 \rangle $ as the partition function of a six-vertex model on the cylinder, with boundary conditions in the imaginary time direction encoded by the initial state $| \Psi_0\rangle$. There are $2N$ horizontal rows, each line corresponding to the action of the transfer matrix $t(u)$, where $u =- \beta/2N, \beta/2N-\eta$ for even/odd rows respectively.}
 \label{fig:partitionfunction1}
 \end{figure}
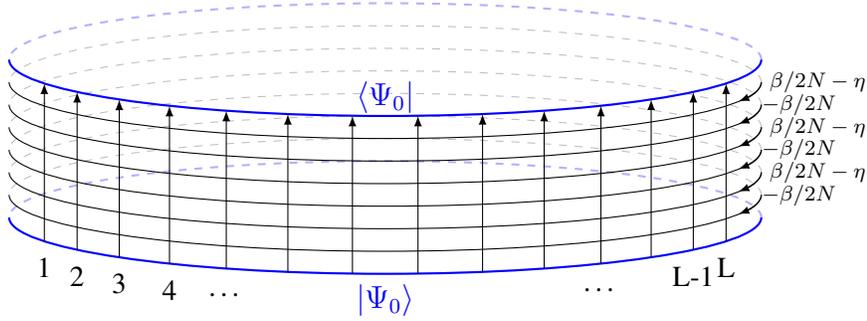

In this section we review the idea behind the computation of the dynamical free energy \eqref{eq:g_en_function} by means of the quantum transfer matrix approach. Following \cite{pozsgay-13}, our starting point is given by the well known Suzuki-Trotter decomposition
\be
e^{-w H }=\lim_{N \to \infty} \left(1 - \frac{w H}{N}\right)^N \,.
\label{eq:suz_trot}
\ee
Next, this expression can be cast in a form more suitable for further analytical investigation. In particular, for large $N$ one can rewrite \cite{klumper-04, pozsgay-13}
\bea 
 \left(1 - \frac{w H}{N}\right)^N \simeq   \frac{   \left[ t(-\beta_w/2N) t(-\eta + \beta_w/2N ) \right]^N }
  {\sinh(-\beta_w/2N + \eta)^{2 L N}}  \,,
  \label{SuzukiTrotter}
\eea
where we defined
\be
\beta_w=\frac{J}{2}\sinh(\eta)w\,,
\label{beta_parameter}
\ee
and where $J$ and $\eta$ are given in \eqref{hamiltonian} and \eqref{eta_parameter}. In the following the dependence on $w$ will be omitted when this will not generate confusion and we will simply write $\beta$ instead of $\beta_{w}$. In \eqref{SuzukiTrotter} we introduced the well-known transfer matrix $t(u)$, which is one of the central objects in the algebraic Bethe ansatz construction to diagonalize the Hamiltonian \eqref{hamiltonian} \cite{kbi-93}. It is defined as 
\be
 t(u) = {\rm tr}_0 \left\{\mathcal{L}_L(u) \ldots \mathcal{L}_1(u)\right\} \,,
 \label{eq:transfermatrixt}
\ee   
where the trace is taken over the auxiliary quantum space $h_{0}\simeq \mathbb{C}^2$. The Lax operators $\mathcal{L}_j(u)$ are written in terms of the $R$-matrix corresponding to the six-vertex model
\be 
\mathcal{L}_j (u) = R_{0,j}(u) \,, 
\label{eq:mathcal_L}
\ee
which in turn can be written as 
\be  
R(u) 
= 
\left(
\begin{array}{cccc}
\sinh(u+\eta) & & & \\
 & \sinh u & \sinh \eta & \\
&\sinh \eta  & \sinh u  & \\
 & & & \sinh(u+\eta)
\end{array}
\right)
\,.
\label{Rmatrix}
\ee
The transfer matrices \eqref{eq:transfermatrixt} for different values of the spectral parameter $u$ commute with one another. Furthermore, Eq.~\eqref{SuzukiTrotter} is an immediate consequence of the following identity \cite{klumper-04}
\be
 \frac{t(-\beta/2N) t(-\eta + \beta/2N )}{\sinh(-\beta/2N + \eta)^{2 L }} 
 = 
 1 - \frac{2\beta}{JN\sinh \eta} H  
 + O\left(\frac{1}{N^2}\right) \,.
\ee

Equations \eqref{eq:suz_trot}, \eqref{SuzukiTrotter} allow to greatly simplify our problem. Indeed, the quantity $ \langle \Psi_0 | \left[ t(-\beta/2N) t(-\eta + \beta/2N ) \right]^N  |  \Psi_0 \rangle $ may now be interpreted as the partition function of a six-vertex model on a square lattice. The latter has $L$ vertical and $2N$ horizontal lines, with periodic boundary conditions in the horizontal (space) direction and boundary conditions specified by $ |  \Psi_0 \rangle $ in the vertical (imaginary time) direction (cf. Fig.~\ref{fig:partitionfunction1}).

\begin{figure}
\centering
 \begin{tikzpicture}
\begin{scope}[rotate=-90]
 \draw[thick, blue!30 , dashed] (0,0.5) arc (0:180:3 and 0.75);
 \draw[gray!50, dashed] (0,0) arc (0:180:3 and 0.75); 
 \draw[ gray!50, dashed] (0,-0.5) arc (0:180:3 and 0.75); 
 \draw[ gray!50, dashed] (0,-1) arc (0:180:3 and 0.75); 
   \draw[ gray!50, dashed] (0,-1.5) arc (0:180:3 and 0.75);
  \draw[gray!50, dashed] (0,-2) arc (0:180:3 and 0.75);
      \draw[ gray!50, dashed] (0,-2.5) arc (0:180:3 and 0.75);
   \draw[thick, blue!30 , dashed] (0,-3) arc (0:180:3 and 0.75);

   \draw[blue, thick] (0,0.5) arc (0:-180:3 and 0.75);
\draw (0,0) arc (0:-180:3 and 0.75);\draw[->,>=latex] (0,0) arc (0:-20:3 and 0.75);  
 \draw (0,-0.5) arc (0:-180:3 and 0.75);\draw[->,>=latex] (0,-0.5) arc (0:-20:3 and 0.75); 
\draw (0,-1) arc (0:-180:3 and 0.75);\draw[->,>=latex] (0,-1) arc (0:-20:3 and 0.75);  
\draw (0,-1.5) arc (0:-180:3 and 0.75);\draw[->,>=latex] (0,-1.5) arc (0:-20:3 and 0.75);  
 \draw (0,-2) arc (0:-180:3 and 0.75);\draw[->,>=latex] (0,-2) arc (0:-20:3 and 0.75);  
   \draw (0,-2.5) arc (0:-180:3 and 0.75);\draw[->,>=latex] (0,-2.5) arc (0:-20:3 and 0.75); 
\draw[blue, thick] (0,-3) arc (0:-180:3 and 0.75); 
   
\node[blue] at   ($(-3,1.25)+(-90:5 and 0.75)$) {$| \Psi_0 \rangle$};
\node[blue] at   ($(-3,-3.5)+(-90:5 and 0.75)$) {$\langle \Psi_0 |$};

\foreach \angle in {-155,-135,...,-25}
{ 
\draw[] ($(-3,-3)+(\angle:3 and 0.75)$) -- ($(-3,0.5)+(\angle:3 and 0.75)$);
\draw[] ($(-3,-3)+(\angle-10:3 and 0.75)$) -- ($(-3,0.5)+(\angle-10:3 and 0.75)$);
}

\node at (0.5,-0.) {\scriptsize $\xi_{1}$};
\node at (0.5,-0.5) {\scriptsize $\xi_{2}$};
\node at (0.5,-1.5) {\scriptsize $\ldots$};
\node at (0.5,-2.5) {\scriptsize $\xi_{2N}$};
\end{scope}

\node at (2.5,2) {\large =};

\begin{scope}[shift={(8,0)}]
\begin{scope}[rotate=-90]
 \draw[gray!50, dashed] (0,0) arc (0:180:3 and 0.75);
\draw[ gray!50, dashed] (0,-0.5) arc (0:180:3 and 0.75);
 \draw[ gray!50, dashed] (0,-1.5) arc (0:180:3 and 0.75);
 \draw[gray!50, dashed] (0,-2) arc (0:180:3 and 0.75);
\draw[ gray!50, dashed] (0,-1) arc (0:180:3 and 0.75);
  \draw[ gray!50, dashed] (0,-2.5) arc (0:180:3 and 0.75);

 \draw (0,0) arc (0:-180:3 and 0.75);\draw[->,>=latex] (0,0) arc (0:-20:3 and 0.75); 
\draw (0,-0.5) arc (0:-180:3 and 0.75);\draw[->,>=latex] (0,-0.5) arc (0:-20:3 and 0.75);  
 \draw (0,-1) arc (0:-180:3 and 0.75);\draw[->,>=latex] (0,-1) arc (0:-20:3 and 0.75);  
 \draw (0,-1.5) arc (0:-180:3 and 0.75);\draw[->,>=latex] (0,-1.5) arc (0:-20:3 and 0.75);  
  \draw (0,-2) arc (0:-180:3 and 0.75);\draw[->,>=latex] (0,-2) arc (0:-20:3 and 0.75);  
 \draw (0,-2.5) arc (0:-180:3 and 0.75);\draw[->,>=latex] (0,-2.5) arc (0:-20:3 and 0.75); 

\foreach \angle in {-155,-145,...,-25}
{ 
}
\foreach \angle in {-155,-135,...,-25}
{ 
\draw[-<,>=latex,rounded corners=4pt] ($(-3,-2.75)+(\angle:3 and 0.75)$)  -- ($(-3,0.5)+(\angle:3 and 0.75)$)--($(-3,0.5)+(\angle-10:3 and 0.75)$) -- ($(-3,-3)+(\angle-10:3 and 0.75)$) -- ($(-3,-3)+(\angle:3 and 0.75)$)  -- ($(-3,-2.75)+(\angle:3 and 0.75)$) ;
\node at ($(-3.,-3.5)+(\angle-5:3 and 0.75)$) {$K^+$};
\node at ($(-3.,1)+(\angle-5:3 and 0.75)$) {$K^-$};
}

\node at (0.5,-0.) {\scriptsize $\xi_{1}$};
\node at (0.5,-0.5) {\scriptsize $\xi_{2}$};
\node at (0.5,-1.5) {\scriptsize $\ldots$};
\node at (0.5,-2.5) {\scriptsize $\xi_{2N}$};
\end{scope}
\end{scope}
\end{tikzpicture}
  \caption{Pictorial representation of the partition function of Fig.~\ref{fig:partitionfunction1} after reflection with respect to the North-West diagonal. Using the factorized form \eqref{eq:initial} of the initial state, the partition function may be viewed as generated by a  quantum transfer matrix associated with the monodromy matrix $T^{\rm QTM}(u)$. The inhomogeneties are $\xi_j = \beta/2N, -\beta/2N + \eta$ for $j$ even/odd respectively and the boundary reflection matrices $K^{\pm}$ encode the dependence on $| \psi_0 \rangle$ [defined in \eqref{eq:initial}] and hence on the initial state.}
 \label{fig:partitionfunction2}
 \end{figure}
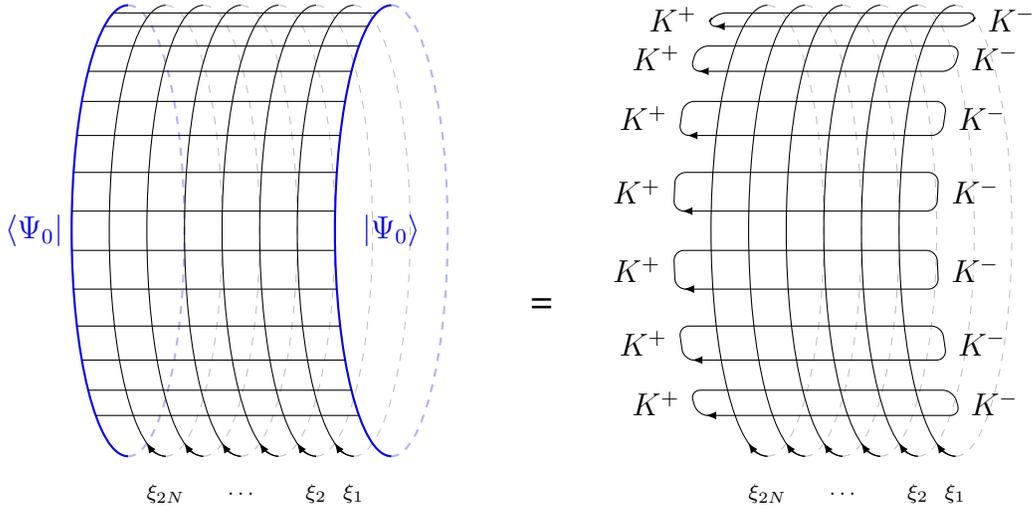

Under a reflection along the North-West diagonal (which leaves the weights of the six-vertex model invariant), and using the factorized structure \eqref{eq:initial} of the initial states considered in this work, the partition function can be reinterpreted as generated from a new transfer matrix. This is the so-called quantum transfer matrix, which acts in the original space direction as pictorially represented in Fig.~\ref{fig:partitionfunction2}, and which is generated from the following monodromy matrix  
\bea
T^{\rm QTM}(u)&=&L_{2N,0}(u-\beta/2N)L_{2N-1,0}(u+\beta/2N-\eta)\cdots\nonumber\\
 &\cdots & L_{2,0}(u-\beta/2N)L_{1,0}(u+\beta/2N-\eta)\,.
\eea
In fact, one can easily derive
\be
\fl \langle \Psi_0 | \left[ t(-\beta/2N) t(-\eta + \beta/2N ) \right]^N  |  \Psi_0 \rangle = {\rm tr}\left\{\left[\langle \psi_0 |T^{\rm QTM}(0)\otimes T^{\rm QTM}(0) |\psi_0\rangle\right]^{L/2}\right\}\,,
\label{aux_eq}
\ee
where $|\psi_0\rangle$ is defined by the initial state through \eqref{eq:initial} and where the trace in the r. h. s. is along the physical spacial direction.
Finally, defining
\be
\mathcal{T}=\frac{\langle \psi_0 |T^{\rm QTM}(0)\otimes T^{\rm QTM}(0) |\psi_0 \rangle}{\left[\sinh(-\beta/2N+\eta)\right]^{4N}}\,,
\label{eq:mathcal_t}
\ee
and putting everything together we arrive at 
\bea 
 \langle \Psi_0 | e^{-w H } |  \Psi_0 \rangle  = 
  \lim_{N \to \infty}{\rm tr}\left[  
  \mathcal{T}^{L/2}\right]\,.
 \,
 \label{loschmidtQTM}
\eea
Following \cite{pozsgay-13}, we call $\mathcal{T}$ the boundary quantum transfer matrix.

In analogy with the thermal case \cite{klumper-04}, we now consider the following two assumptions
\begin{itemize}
\item For real values of the parameter $w$, the boundary quantum transfer matrix $\mathcal{T}$ has a leading eigenvalue $\Lambda_0$ whose absolute value remains separated from that of the subleading eigenvalues by a finite gap, even in the $N\to\infty$ limit. 
\item The large $L$ behaviour of (\ref{loschmidtQTM}) can be studied by exchanging the limits $N \to \infty$ and $L \to \infty$. 
\end{itemize} 
If these assumptions are verified, it is straightforward to obtain in the large $L$ limit 
\bea 
 \langle \Psi_0 | e^{-w H } |  \Psi_0 \rangle  \simeq 
  \left(\lim_{N \to \infty}  
 {\Lambda_0}\right)^{L/2}
 \,,
 \label{loschmidtLambda}
\eea
and hence
\be
g(w) = \lim_{L \to \infty} \frac{1}{L} \log \langle \Psi_0 | e^{-w H } |  \Psi_0 \rangle  
= \frac{1}{2} \lim_{N\to \infty} \log \Lambda_0  \,.
\label{gLambda}
\ee 
This formula is the starting point for the analytical derivation of the dynamical free energy \eqref{eq:g_en_function}. Indeed, we are now left with the problem of computing the leading eigenvalue of the boundary quantum transfer matrix $\mathcal{T}$.

As we already mentioned, the form of $\mathcal{T}$ explicitly depends on the initial state considered. In the case of the N\'eel state \eqref{eq:neel} the computation of $\Lambda_0$ was performed in \cite{pozsgay-13}, where $\mathcal{T}$ was diagonalized by means of the so called {\it diagonal} boundary algebraic Bethe ansatz and the Trotter limit computed. As we will see in the next section, in the case of more general initial states of the form \eqref{eq:initial}, one needs to resort to the {\it non-diagonal} version of the boundary algebraic Bethe ansatz and additional difficulties arise.

In the next section we first review the diagonal case corresponding to the N\'eel state and later discuss the more general non-diagonal boundary algebraic Bethe ansatz. These results will then be used in section~\ref{sec:TBA_equations} where an approach for the computation of the Trotter limit  different to the one of \cite{pozsgay-13} is proposed. The latter is based on the derivation of non-linear integral equations from the so called fusion of boundary transfer matrices. As we will see, one of the advantages of this method is that it can be straightforwardly applied both in the diagonal and non-diagonal cases. 

\section{The boundary algebraic Bethe ansatz: diagonal and non-diagonal boundaries}\label{sec:boundary_bethe}

The boundary algebraic Bethe ansatz is an analytical method which allows to diagonalize Hamiltonians of open spin chains with integrable boundary conditions \cite{skly-88}. Here we only review the aspects relevant to our work, while we refer to the specialized literature for a more systematic treatment \cite{kkmn-07,kmn-14,wycs-15}. 

Given a chain of length $2N$, the construction starts by introducing a boundary transfer matrix $T(u)$ defined as
 \be
T(u) = {\rm tr}_0 \{ K^+(u) T_1(u) K^{-}(u) T_2(u) \} \,.
\label{QTMdef}
\ee
Here we introduced
\bea
T_1(u)&=&\tilde{\mathcal{L}}_{2N}(u)\ldots \tilde{\mathcal{L}}_1(u)\,,\\
\tilde{\mathcal{L}}_j(u)&=&R_{0,j}(u-\xi_j)\,,\\
T_2(u)&=&(-1)^{2N}\sigma_0^yT_1^{t_0}(-u)\sigma_0^y\,=\sigma_0^yT_1^{t_0}(-u)\sigma_0^y\,\label{eq:t2_matrix} \nonumber\\
&=&R_{1,0}(u+\xi_1-\eta)\ldots R_{2N,0}(u+\xi_{2N}-\eta)\,,
\eea
where the Pauli matrix $\sigma_0^y$ acts on the auxiliary space $h_0\simeq \mathbb{C}^{2}$ and where $T_1^{t_0}$ indicates transposition in $h_0$. The last equality follows from the properties of the $R$-matrix defined in \eqref{Rmatrix} \cite{kkmn-07}, while the inhomogeneities $\xi_j$ are parameters which  for the moment are left arbitrary. Finally,  the trace in \eqref{QTMdef} is performed over the auxiliary space $h_0$ [not to be confused with the auxiliary space of the physical transfer matrix \eqref{eq:transfermatrixt}].

\begin{figure}
\centering
 \begin{tikzpicture}
\draw[rounded corners=10pt] (0.5,0) -- (8,0) -- (8.5,0.5) -- (8,1) -- (0,1) -- (-0.5,0.5) -- (0,0) -- (0.5,0);
\node at (-1,0.5) {$K^+(u)$};
\node at (9,0.5) {$K^-(u)$};

\foreach \x in {0.5,1.5,...,7.5} { 
\draw[>-,>=latex] (\x,-0.25)  -- (\x,1.25);
}
\node at (0.5,-0.5) {\small $\xi_{2N}$};
\node at (4,-0.5) {\small $\ldots$};
\node at (6.5,-0.5) {\small $\xi_{2}$};
\node at (7.5,-0.5) {\small $\xi_{1}$};
\end{tikzpicture}
  \caption{Symbolic representation of the transfer matrix $T(u)$ in Eq. (\ref{QTMdef}), acting on $2N$ sites with inhomogeneous spectral parameters $\xi_j$.}
 \label{fig:QTM}
 \end{figure}
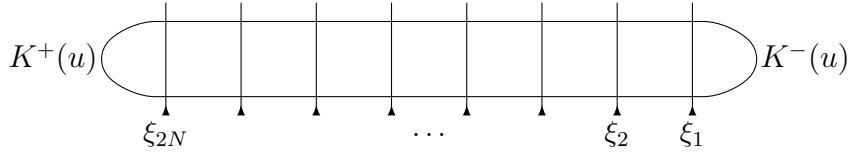

The boundary reflection matrices $K^{\pm }(u)$ are $2\times 2$ matrices 
\be
K^{\pm}(u) = \left(\begin{array}{cc}
k^\pm_{11}(u)&k^\pm_{12}(u) \\
k^\pm_{21}(u)&k^\pm_{22}(u)
\end{array}\right) \,,
\label{eq:k_matrices}
\ee
which are solution of the so called reflection equations \cite{skly-88}. The boundary transfer matrix \eqref{QTMdef} is symbolically represented in Fig.~\ref{fig:QTM}. 

The relevance of this construction for our purposes lies in the possibility of interpreting the boundary quantum transfer matrix $\mathcal{T}$ in \eqref{eq:mathcal_t} as an operator of the form \eqref{QTMdef}. This in turn allows us to employ boundary algebraic Bethe ansatz techniques for the computation of the leading eigenvalue $\Lambda_0$. We explicitly show this in the following.

First, introducing the components of $T_1(u)$ in the auxiliary space $h_0$ as
\be
T_1(u) = \left( 
\begin{array}{cc}
A(u) & B(u) \\ 
C(u) & D(u)  
\end{array}
\right) \,,
\label{eq:t1_components}
\ee 
it follows from \eqref{eq:t2_matrix} that
\be
T_2(u) = \left( 
\begin{array}{cc}
D(-u) & -B(-u) \\ 
-C(-u) & A(-u)  
\end{array}
\right) \,,
\label{eq:t2_components}
\ee 
where the components $A(u)$, $B(u)$, $C(u)$, $D(u)$ are operators acting on the physical space $(\mathbb{C}^2)^{\otimes{2N}}$. Using \eqref{eq:t1_components}, \eqref{eq:t2_components}, it is now straightforward to rewrite $T(u)$ in \eqref{QTMdef} as
\be
T(u)=\langle v^{+}(u)|T_1(u)\otimes T_1(-u)|v^{-}(u)\rangle\,,
\label{eq:final_form}
\ee
where we introduced the vectors $|v^{\pm}(u)\rangle$ defined as
\bea
 |v^-(u)\rangle=- k_{12}^-(u)|\uparrow\uparrow\rangle+k_{11}^-(u)|\uparrow\downarrow\rangle-k_{22}^-(u)|\downarrow\uparrow\rangle+k_{21}^-(u)|\downarrow\downarrow\rangle\,,\label{v1}\\
 \left(|v^+(u)\rangle\right)^*= -k_{21}^+(u)|\uparrow\uparrow\rangle+k_{11}^+(u)|\uparrow\downarrow\rangle-k_{22}^+(u)|\downarrow\uparrow\rangle+k_{12}^+(u)|\downarrow\downarrow\rangle\,.
\label{v2}
\eea

It is now evident from \eqref{eq:final_form} that $T(0)$ is proportional to $\mathcal{T}$ in \eqref{eq:mathcal_t} provided that
\bea 
\langle v^+(0) | & \propto & \langle \psi_0|   \nonumber \\
| v^-(0) \rangle &  \propto  & | \psi_0 \rangle  \,,
\label{conditionv}
\eea
and that the inhomogeneous spectral parameters are chosen as 
 \bea
\xi_{2j+1}&=&\beta/2N\,,\\
\xi_{2j}&=&\eta-\beta/2N\,.
\label{inhomogeneities}
\eea 
If these conditions are met, one simply obtains
\bea 
\mathcal{T}=   \frac{  1 }
  {\sinh(-\beta/2N + \eta)^{4 N}}  
  \frac{1}{\langle v^+ (0) | v^- (0)\rangle }
 T(0) \,.
 \label{BQTMdef}
 \eea 
This relation allows us to directly obtain the eigenvalues of $\mathcal{T}$ once the eigenvalues of $T(0)$ are known. Note once again that $T(u)$ depends explicitly on the initial state through \eqref{conditionv}. In particular, the identification \eqref{conditionv} fixes the $K$-matrix \eqref{eq:k_matrices} through \eqref{v1}, \eqref{v2}. Different initial states then require diagonal or non-diagonal $K$-matrices. In turn, this makes it necessary to resort to either diagonal or non-diagonal boundary algebraic Bethe ansatz techniques to obtain the eigenvalues of $T(0)$. We now separate the discussion for these two different cases.

\subsection{Diagonal reflection matrices: the N\'eel state}
\label{sec:NeelQTMreview}

In the simplest case, the identification \eqref{conditionv} leads to diagonal $K$-matrices. This is what happens for the N\'eel state \eqref{eq:neel}, which was explicitly considered in \cite{pozsgay-13} (together with the so called Majumdar-Ghosh sate).

A diagonal solution of the reflection equation can be obtained as \cite{kkmn-07} 
\bea
K^{\pm}(u)&=&K(u\pm \eta/2,\xi_{\pm})\,\\
K(u,\xi)&=&
\left(\begin{array}{cc}
\sinh\left(\xi+u\right)& 0 \\
0&\sinh(\xi-u)
\end{array}\right)\label{k_matrix_diag}\,.
\eea
Then, from \eqref{v1}, \eqref{v2}, one can easily see that condition \eqref{conditionv} is satisfied by choosing the boundary parameters $\xi_\pm$ as 
\be
\xi_\pm = \mp \eta/2 \,.
\label{eq:choice}
\ee
With this choice, \eqref{v1} and \eqref{v2} yield
\bea
|v^-(0)\rangle&=&-\sinh(\eta)|\downarrow\uparrow\rangle\,,\\
(|v^+(0)\rangle)^\ast &=&\sinh(\eta)|\downarrow\uparrow\rangle\,.
\eea
The choice \eqref{eq:choice} completely specifies the diagonal $K$-matrix \eqref{k_matrix_diag} and hence the boundary transfer matrix \eqref{QTMdef}, which can then be diagonalized.

The eigenvalues of the transfer matrix $T(u)$, and therefore the leading eigenvalue $\Lambda_0$ of $\mathcal{T}$, can be constructed through the diagonal boundary algebraic Bethe ansatz procedure, which we briefly review here.

Introducing the notation 
\be
{U}_{-}(u) = T_1(u) K^{-}(u) T_2(u) = \left( \begin{array}{cc}
{A}_-(u) &{B}_-(u) \\ 
{C}_-(u) & {D}_-(u)  
\end{array}   \right)    \,,
\ee
the common eigenstates of the operators $T(u)$ are obtained from the ferromagnetic reference eigenstate $|\uparrow \uparrow \ldots \rangle$ as 
\be
| \{ \lambda_j \}_{j=1}^R \rangle = \prod_{j=1}^R
 B_-(\lambda_j) | \uparrow \uparrow \ldots  \rangle   \,.
\ee
Here, the complex parameters  $\lambda_j$, the so-called rapidities, have to be chosen to satisfy the Bethe equations 
\bea
\hspace{-3cm}
\left[ 
\frac{\sinh(\lambda_j + \beta/2N -\eta ) \sinh(\lambda_j - \beta/2N)}{\sinh(\lambda_j - \beta/2N + \eta ) \sinh(\lambda_j + \beta/2N)}
\right]^{2N} 
\prod_{k\neq j}^R \frac{\sinh(\lambda_j - \lambda_k + \eta)\sinh(\lambda_j + \lambda_k + \eta)}{\sinh(\lambda_j - \lambda_k - \eta)\sinh(\lambda_j + \lambda_k - \eta)}
\nonumber \\
\times \frac{\sinh(\lambda_j - (\xi_+- \eta/2))\sinh(\lambda_j - (\xi_- - \eta/2))}{\sinh(\lambda_j +(\xi_+- \eta/2))\sinh(\lambda_j + (\xi_- - \eta/2))}
=
1\,.
\eea
As we will comment later, the leading eigenvalue corresponds to a set of $R=N$ rapidities. In the following we will then restrict to this case.

Given the set $\{\lambda_j\}_{j=1}^N$, and following \cite{pozsgay-13}, it is convenient to introduce the doubled set
\be
\{\tilde{\lambda}_k\}_{k=1}^{2N}=\{\lambda_k\}_{k=1}^{N}\cup\{-\lambda_k\}_{k=1}^{N}\,.
\ee
Defining further
\bea
Q(u)& \equiv &\prod_{k=1}^{2N}\sinh(u-\tilde{\lambda}_k)\,,\\
\phi(u) & \equiv & \prod_{k=1}^{2N}\sinh\left(u-\eta/2+\xi_k\right)\sinh\left(u+\eta/2-\xi_k\right)\,,\label{phi_function}\\
\omega_1(u)&=&\frac{\sinh(2u+\eta)\sinh(u+\xi^+-\eta/2)\sinh(u+\xi^--\eta/2)}{\sinh(2u)}\,,\label{omega1_function}\\
\omega_2(u)&=&\frac{\sinh(2u-\eta)\sinh(u-\xi^++\eta/2)\sinh(u-\xi^-+\eta/2)}{\sinh(2u)}\,,\label{omega2_function}
\eea
the Bethe equations can be rewritten as 
\be
\frac{\omega_2(\lambda_j)}{\omega_1(\lambda_j)}\frac{Q(\lambda_j+\eta)\phi(\lambda_j-\eta/2)}{Q(\lambda_j-\eta)\phi(\lambda_j+\eta/2)}=-1\,.
\label{eq:new_bethe_eq_diagonal}
\ee

A given solution $\bm{\lambda} \equiv \{\lambda_j\}$ of the Bethe equations \eqref{eq:new_bethe_eq_diagonal} corresponds to an eigenvalue of $T(u)$, which we indicate as $T_{\bm{\lambda}}(u)$. It reads
\be
T_{\bm{\lambda}}(u)= \omega_1(u)\phi(u+\eta/2)\frac{Q(u-\eta)}{Q(u)}+\omega_2(u)\phi(u-\eta/2)\frac{Q(u+\eta)}{Q(u)}\,,
\label{TQrelation_diagonal}
\ee
where the dependence on $\{\lambda_j\}$ is encoded in the functions $Q(u)$. Eq.~\eqref{TQrelation_diagonal} is sometimes referred to as the $T-Q$ relation. 

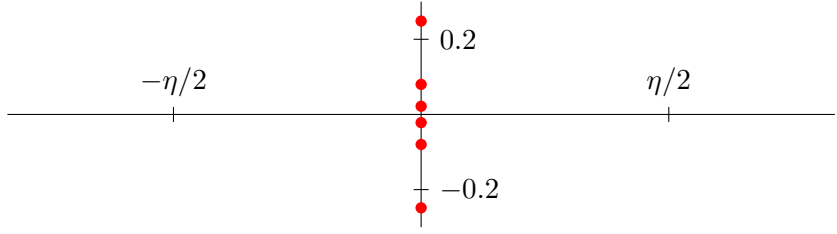
\begin{figure}
\centering
\begin{tikzpicture}[scale=5]

\draw (-1.1,0) -- (1.1,0); 
\draw (0,-0.3) -- (0,0.3); 
\draw (-0.02,0.2) -- (0.02,0.2) node[right] {\small $0.2$};
\draw (-0.02,-0.2) -- (0.02,-0.2) node[right] {\small $-0.2 $};
\draw (0.658479,-0.02) -- (0.658479,0.02) node[above] {\small $\eta/2$};
\draw (-0.658479,-0.02) -- (-0.658479,0.02) node[above] {\small $-\eta/2$};

\foreach \x in {(0.,-0.248486),(0.,+0.248486),(0.,-0.0799621),(0.,+0.0799621),(0.,-0.0218245),(0.,+0.0218245)}
\fill[red] \x circle[radius=0.015];
\end{tikzpicture}
  \caption{Bethe roots $\{\tilde{\lambda}_j\}_{j=1}^{2N}$, as obtained by solving \eqref{eq:new_bethe_eq_diagonal}, corresponding to the leading eigenvalue of the transfer matrix in the diagonal case. The plot corresponds to $2N=6$, $\Delta=2$, $\beta=0.5$. We see that the Bethe roots are located symmetrically along the imaginary axis.}
 \label{fig:rootsdiag}
 \end{figure}

The formal relation (\ref{TQrelation_diagonal}) may be compared with exact diagonalization for finite system sizes $2N$, allowing to find the set of roots $\{\lambda_j \}_{j=1}^{N}$ associated with the eigenstates of interest. By numerical diagonalization of the transfer matrix for $2N=4,6,8,10$, we find that the (unique) leading eigenvalue of $T(0)$ is obtained from a set of $N$ roots $\{\lambda_j \}_{j=1}^N$ which are situated on the imaginary axis. Therefore the rapidities of the corresponding doubled set $\{\tilde{\lambda_j} \}_{j=1}^{2N}$ are distributed symmetrically on the imaginary axis. An example is shown in Fig.~\ref{fig:rootsdiag}. 

Before discussing the Trotter limit of the leading eigenvalue, we present in the next section the case of non-diagonal reflection matrices. The Trotter limit will then be discussed in section~\ref{sec:TBA_equations}, where we employ an approach which can be straightforwardly applied both in the diagonal and non-diagonal cases.

\subsection{Non-diagonal reflection matrices: tilted N\'eel and tilted ferromagnet states}
\label{sec:tNeelQTM}

For more general initial states, the identification \eqref{conditionv} leads through \eqref{v1}, \eqref{v2} to non-diagonal boundary transfer matrices. Let us introduce the general non-diagonal solution of Sklyanin's reflection equation \cite{skly-88,kmn-14,wycs-15}
\bea
K^{\pm}(u)&=&K(u\pm \eta/2,\xi_{\pm},\kappa_{\pm},\tau_{\pm})\,\label{aux_k_matrix}\\
K(u,\xi,\kappa,\tau)&=&
\left(\begin{array}{cc}
\sinh\left(\xi+u\right)&\kappa e^{\tau} \sinh\left(2u\right) \\
\kappa e^{-\tau}\sinh\left(2u\right)&\sinh(\xi-u)
\end{array}\right)\label{k_matrix_nondiag}\,. 
\eea
For later convenience we introduce $\alpha_\pm$, $\beta_\pm$, defined by the following parametrization
\be
\sinh\alpha_{\pm}\cosh\beta_{\pm}=\frac{\sinh\xi_{\pm}}{2\kappa_{\pm}}\,,\quad \cosh\alpha_{\pm}\sinh\beta_{\pm}=\frac{\cosh\xi_{\pm}}{2\kappa_{\pm}}\,.
\label{eq:parametrization}
\ee

Analogously to the diagonal case discussed in the previous section, the parameters $\alpha_\pm$, $\beta_\pm$,  $\tau_{\pm}$ [and hence $\xi_{\pm}$, $\kappa_{\pm}$, through \eqref{eq:parametrization}] can be chosen in such a way that the condition \eqref{conditionv} is satisfied for a given initial state. In the following we report the choice of the parameters for tilted N\'eel and tilted ferromagnet states.

\begin{itemize}
\item{{\it Tilted N\'eel state}. In this case, it is straightforward to verify that the parameters of the $K$-matrix can be chosen as
\bea
\tau_\pm &=&0\,,\label{par_1}\\
\alpha^{\mp}&=&\pm\eta/2\,,\label{par_2}\\
\beta_{\pm}&=&\zeta\,,\label{par_3}\\
e^{-\zeta}&=&\tan\left(\frac{\vartheta}{2}\right) \label{def_zeta}\,,
\eea
where $\vartheta$ is the tilting angle in the definition \eqref{tilted_neel}. With this choice condition \eqref{conditionv} is met. Explicitly,
\be
-(|v^+(0)\rangle)^*=|v^-(0)\rangle=-\frac{\kappa\sinh(\eta)}{\sin(\vartheta/2)\cos(\vartheta/2)}|\vartheta;\swarrow\nearrow\rangle\,\,.
\ee
Note in particular that if $\vartheta \to 0$ then $\zeta \to \infty$, and using $\kappa\sim e^{-\zeta}$ one consistently recovers the result of the previous section for the N\'{e}el state. 
}
\item{{\it Tilted ferromagnet}. The parameters of the $K$-matrix are now chosen as
\bea
\alpha^{\mp}&=&\pm\eta/2\,,\label{fpar1}\\
\beta_{\pm}&=&i\frac{\pi}{2},\\
\tau_\pm &=&\pm \left(i\frac{\pi}{2}+r\right)\,,\\
e^{-r}&=&{\rm cotan}\left(\frac{\vartheta}{2}\right)\,,\label{fpar4}
\eea
where again $\vartheta$ is the tilting angle in the definition \eqref{tilted_ferromagnet}. Using this choice, it is straightforward to verify
\be
-(|v^+(0)\rangle)^*=|v^-(0)\rangle=-i\frac{\kappa\sinh(\eta)}{\cos(\vartheta/2)\sin(\vartheta/2)}|\vartheta;\nearrow\nearrow\rangle\,.
\ee
}

\end{itemize}

After fixing the parameters of the $K$-matrix \eqref{k_matrix_nondiag}, the spectral problem associated with the boundary transfer matrix \eqref{QTMdef} can be addressed. The procedure described in section \ref{sec:NeelQTMreview} for the diagonal case cannot be applied directly here. This is because the ferromagnetic state $| \uparrow \uparrow \ldots  \rangle $ is not anymore an eigenstate of the transfer matrix $T(u)$, and cannot therefore be used as the reference state for the construction of all eigenstates.  

In fact, the long-standing problem of completely characterizing the spectrum of the boundary transfer matrix for arbitrary $K$-matrices has only recently been solved, as a result of the combined effort of several groups \cite{kmn-14,wycs-15, nepomechie-02,clsw-03,fgsw-11,niccoli-12,cysw-13,nepomechie-13} (see \cite{kmn-14,wycs-15} for some historical details on these interesting developments). In particular, an important discovery was that a generalized version of the $T-Q$ relation \eqref{TQrelation_diagonal} could be recovered also in the non-diagonal case, involving again a finite set of rapidities. In turn, these can be obtained as the solution of an appropriate set of Bethe equations. Here we simply report the results which are directly relevant to our work, while we refer to \cite{kmn-14,wycs-15} for a thorough treatment.

In the following we employ many of the notations used in \cite{kmn-14}. We start by introducing the so called inhomogeneous $T-Q$ relation verified by the eigenvalues $T_{\bm{\lambda}}(u)$ of the boundary transfer matrix, which is written as
\be
\frac{T_{\bm{\lambda}}(u)}{\sinh(\xi_+)\sinh(\xi_-)}={\bf A}(u)\frac{Q(u-\eta)}{Q(u)}+{\bf A}(-u)\frac{Q(u+\eta)}{Q(u)}+\frac{F(u)}{Q(u)}\,.
\label{TQrelation_nondiagonal}
\ee
We will now define the functions appearing above. First, the parameter $\xi_{\pm}$ are defined by the $K$-matrix \eqref{aux_k_matrix}, while
\be
F(u)=2^{4N}F_0\left(\cosh^2(2u)-\cosh^2\eta\right)
\phi\left( u + \frac{\eta}{2} \right)\phi\left( u - \frac{\eta}{2} \right)
\,,
\ee
where $\phi(u)$ was introduced in \eqref{phi_function} and where
\be
\fl F_0=\frac{2\kappa_{+}\kappa_{-}\left(\cosh(\tau_{+}-\tau_{-})-\cosh(\alpha_{+}+\alpha_{-}-\beta_++\beta_--(2N+1)\eta)\right)}{\sinh\xi_+\sinh\xi_-}\,.
\ee
Here the parameters $\tau_{\pm}$, $\alpha_{\pm}$, $\beta_\pm$, and $\kappa_\pm$ are  defined in \eqref{k_matrix_nondiag} and \eqref{eq:parametrization}. Further, the function ${\bf A}(u)$ is given by
\be
{\bf A}(u)=(-1)^{2N}\frac{\sinh(2u+\eta)}{\sinh(2u)}g_{+}(u)g_{-}(u)  \phi\left( u + \frac{\eta}{2} \right) \,,
\label{eq:bold_a_function}
\ee
where 
\be
g_{\pm}(u)=\frac{\sinh(u+\alpha_{\pm}-\eta/2)\cosh(u\mp\beta_{\pm}-\eta/2)}{\sinh\alpha_{\pm}\cosh\beta_{\pm}}\,.
\ee
Finally, the $Q$-functions are parametrized by a set of rapidities $\{\lambda_j\}_{j=1}^{2N}$ as
\be
Q(u)=2^{2N}\prod_{j=1}^{2N}\left(\cosh2u-\cosh2\lambda_j\right)\,,
\label{q_functions_aux}
\ee
which can be rewritten, introducing analogously to the diagonal case the doubled set $\{\tilde{\lambda}_j\}_{j=1}^{4N}=\{\lambda_j\}_{j=1}^{2N}\cup\{-\lambda_j\}_{j=1}^{2N}$, as
\be
Q(u)=2^{4N}\prod_{j=1}^{4N}\sinh\left(u-\tilde{\lambda}_j\right)\,.
\label{q_functions}
\ee
In this case the (doubled) set of rapidities $\{\tilde{\lambda}_j\}_{j=1}^{4N}$ is determined as the solution of a new set of Bethe equations containing an additional term, namely
\be
{\bf A}(\lambda_k)Q(\lambda_k-\eta)+{\bf A}(-\lambda_k)Q(\lambda_k+\eta)=-F(\lambda_k)\,.
\label{eq:inh_bethe_eq}
\ee
We stress that the inhomogeneous $T - Q$ relation \eqref{TQrelation_nondiagonal} has to be understood as follows: for every eigenstate of the boundary transfer matrix \eqref{QTMdef}, there exist a set of solutions $\bm{\lambda} \equiv \{\lambda_j\}_{j=1}^{2N}$ of the inhomogeneous Bethe equations \eqref{eq:inh_bethe_eq} for which the corresponding eigenvalue can be written as \eqref{TQrelation_nondiagonal}.
Note that in the diagonal case the number $R$ of Bethe roots $\{\lambda_j \}_{j=1}^R$ is fixed by the value of the total magnetization (which commutes with the transfer matrix and is therefore well defined for each eigenstate). Here, instead the latter is not conserved and the number of Bethe roots is always exactly equal to the number of sites of the chain, namely $2N$.

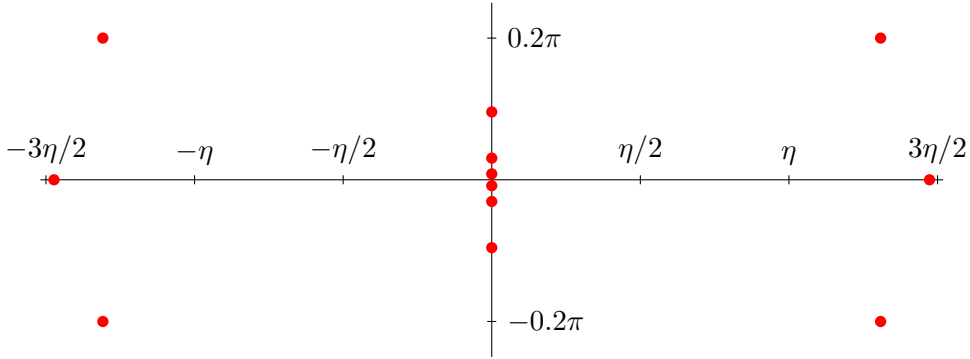
\begin{figure}
\centering
\begin{tikzpicture}[scale=3]

\draw (-2,0) -- (2,0); 
\draw (0,-0.785398) -- (0,0.785398); 
\draw (-0.02,0.628319) -- (0.02,0.628319) node[right] {\small $0.2 \pi$};
\draw (-0.02,-0.628319) -- (0.02,-0.628319) node[right] {\small $-0.2 \pi$};
\draw (0.658479,-0.02) -- (0.658479,0.02) node[above] {\small $\eta/2$};
\draw (1.31696,-0.02) -- (1.31696,0.02) node[above] {\small $\eta$};
\draw (1.97544,-0.02) -- (1.97544,0.02) node[above] {\small $3\eta/2$};
\draw (-0.658479,-0.02) -- (-0.658479,0.02) node[above] {\small $-\eta/2$};
\draw (-1.31696,-0.02) -- (-1.31696,0.02) node[above] {\small $-\eta$};
\draw (-1.97544,-0.02) -- (-1.97544,0.02) node[above] {\small $-3\eta/2$};

\foreach \x in {(-1.72327,-0.628511),(-1.72327,0.628511),(-1.93988,0),(0.,-0.30092),(0.,0.30092),(0.,-0.0961014),(0.,0.0961014),(0.,-0.0262106),(0.,0.0262106),(1.72327,-0.628511),(1.72327,0.628511),(1.93988,0)}
\fill[red] \x circle[radius=0.025];
\end{tikzpicture}
  \caption{Bethe roots $\{\tilde{\lambda}_j\}_{j=1}^{4N}$, as obtained by solving \eqref{eq:inh_bethe_eq}, associated with the leading eigenvalue of the transfer matrix in the non-diagonal case. The plot corresponds to $2N=6$, $\Delta= 4$, $\beta=0.2$, and boundary parameters associated with the tilted N\'eel state at $\vartheta=\frac{\pi}{3}$.  We recognize the structure discussed in section~\ref{sec:tNeelQTM} with $2N$ roots situated along the imaginary axis ($\lambda_j^{\rm reg}$) and $2N$ additional roots with non-zero real part, located symmetrically with respect to the imaginary axis ($ \lambda_j^{\rm extra}$).}
 \label{fig:rootsnondiag}
 \end{figure}

As in the diagonal case, we can compare the formal relation (\ref{TQrelation_nondiagonal}) with exact diagonalization for small system sizes of length $2N=2,4,6,8$ and we observe that there is once again a unique leading eigenvalue. In the next sections, we mainly focus on the case of tilted N\'eel states, for which we provide explicit results for the dynamical free energy. In this case we studied in detail the corresponding doubled set of Bethe roots $\{\tilde{\lambda}_j\}_{j=1}^{4N}$, which display a recognizable structure. In particular, we observed that they organize into two disjoint sets as
\be
\{\tilde{\lambda}_j\}_{j=1}^{4N}=\{\tilde{\lambda}^{\rm reg}_j\}_{j=1}^{2N}\cup\{\tilde{\lambda}^{\rm extra}_j\}_{j=1}^{2N}\,, 
\ee
where
\begin{itemize}
\item the $2N$ roots $\tilde{\lambda}_j^{\rm reg}$ are displaced symmetrically along the imaginary axis\,,
\item the $2N$ roots $\tilde{\lambda}_j^{\rm extra}$ are distributed as pairs of common imaginary part and opposite real parts.  
\end{itemize}
We report an example of this structure in the complex plane in figure \ref{fig:rootsnondiag}.

For the tilted N\'eel state one can study the diagonal limit $\vartheta \to 0$. In this limit, the real parts of the roots $\tilde{\lambda}_j^{\rm extra}$ diverge proportionally to $\pm \zeta$ [defined in \eqref{def_zeta}]. Their contributions to ratios of $Q$ functions cancel out in the $T-Q$ relation \eqref{TQrelation_nondiagonal}, while the inhomogeneous $F$ term vanishes. In this limit, one can then factor the contributions of these extra roots from the $Q$ function, and the roots $\tilde{\lambda}_j^{\rm reg}$ satisfy the homogeneous relation (\ref{TQrelation_diagonal}). In fact we observed that even for finite nonzero $\vartheta $ the values of the roots $ \tilde{\lambda}_j^{\rm reg}$ are close to the solutions of the corresponding homogeneous $T-Q$ system at $\vartheta=0$.

The results of this and the previous sections give access to the leading eigenvalue of the transfer matrix \eqref{QTMdef} for small Trotter number $N$. In the next section we address the problem of computing the Trotter limit $N\to\infty$ which directly yields the dynamical free energy \eqref{eq:g_en_function}.

\section{Integral equations from fusion of boundary transfer matrices}
\label{sec:TBA_equations}

We now finally address the Trotter limit \eqref{gLambda} required in
the computation of the dynamical free energy
\eqref{eq:g_en_function}. In \cite{pozsgay-13} this problem was solved
for the diagonal case by introducing a single non-linear integral equation for an auxiliary function in the complex plane which could be directly related to the leading eigenvalue $\Lambda_0$ of the transfer matrix \eqref{eq:mathcal_t}.
The method employed in \cite{pozsgay-13} heavily relied on the
particular structure of the Bethe roots in the complex plane. As we
already stressed, the latter is very simple in the diagonal case, as
one can immediately see in Fig.~\ref{fig:rootsdiag}. By contrast, in
the non-diagonal case, the picture is significantly more involved due
to the presence of the additional Bethe roots $\tilde{\lambda}^{\rm
  extra}_j$ as discussed in section~\ref{sec:tNeelQTM} (cf. also
Fig.~\ref{fig:rootsnondiag}). As a consequence, the approach used in
\cite{pozsgay-13} can not be directly applied and a more sophisticated
analysis is required to adapt it to the non-diagonal case\footnote{
In the closely related problem of the physical spin chain with
transverse boundary magnetic fields similar complications occur, and 
the problem of finding the Bethe roots and considering the
  thermodynamic limit of their distribution is not yet settled \cite{Nepo}. }.

In this work we consider a different approach based on the derivation
of non-linear integral equations from fusion of boundary transfer
matrices. This procedure has been employed many times in the thermal case, where it is now well established \cite{kp-92,tsk-01}. On the one hand, this method can be applied directly both in
the diagonal and non-diagonal cases considered in this work. On the other hand, even in the
diagonal case discussed in \cite{pozsgay-13} it is seen to be more
stable when continuation to real times is addressed as discussed
in section~\ref{sec:resolution}. The main qualitative difference is
that the final result is written in terms of an infinite set of
non-linear integral equations, as opposed to the single non-linear
integral equation derived in \cite{pozsgay-13}.  

As a preliminary step, we introduce in the next subsection the family
of fused boundary transfer matrices which can be build out of
\eqref{QTMdef}. As we will explain in the following, the main idea is
to relate the leading eigenvalue of $\mathcal{T}$ in
\eqref{eq:mathcal_t} to the solution of the $T$-system of fused
boundary transfer matrices. In turn, this can be obtained from the
corresponding $Y$-system, which can be conveniently cast in the form
of partially decoupled integral equations. The rest of this section is
devoted to following this program, and each step will be explained in
full detail. 

\subsection{The $T-$system and $Y-$system for boundary quantum transfer matrices}\label{t_y_system}

It is an established result that the boundary transfer matrix $T(u)$ in \eqref{QTMdef} can be used to build an infinite family of transfer matrices $\{T^{(n)}(u)\}_{n=0}^{\infty}$ by the so called fusion procedure \cite{zhou-95,mn-92,zhou-96}, in complete analogy with the case of periodic boundary conditions.

The fused transfer matrices $T^{(n)}(u)$ act on the same space as the transfer matrix $T(u)$ and form a commuting set, namely
\be
\left[T^{(j)}(u),T^{(k)}(w)\right]=0\,,\qquad j,k=0,1,\ldots\,,
\label{eq:commutation}
\ee
with the further identification
\bea
T^{(0)}(u)&\equiv & 1\,,\nonumber\\
T^{(1)}(u)&=& T(u)\label{eq:first:t}\,.
\eea
The family of fused boundary transfer matrices satisfy a set of functional relations known as the $T$-system \cite{zhou-96} 
\bea
T^{(n)}\left(u+\eta/2\right)T^{(n)}\left(u-\eta/2\right)&=&T^{(n-1)}(u)T^{(n+1)}(u)+\Phi_n(u)\,,
\label{t_system}
\eea
where
\bea
\Phi_n(u) =\prod_{j=1}^{n}f\left(u-(n+2-2j)\eta/2 \right)\,.
\eea
The function $f(u)$ encodes the information about the boundary reflection $K$-matrices \eqref{eq:k_matrices}. In the general non-diagonal case, it reads \cite{zhou-96}
\bea
f(u)&=&\frac{\Omega_{+}(u)\Omega_{-}(u)}{\sinh(2u)\sinh(2u+2\eta)}
\phi\left( u + \frac{3\eta}{2} \right)\phi\left( u - \frac{\eta}{2} \right)
\label{f_function}
\eea
where the function $\phi(u)$ is defined in (\ref{phi_function}), while
\bea
\Omega_{+}(u)&=&\sinh(2u+3\eta)\left\{\sinh\left(\xi^+-u-\frac{\eta}{2}\right)\sinh\left(\xi^++u+\frac{\eta}{2}\right)\right.\nonumber\\
&-& \left. (\kappa_+)^2\sinh^2(2u+\eta)\right\}\,,\\
\Omega_{-}(u)&=&\sinh(2u-\eta)\left\{\sinh\left(\xi^-+u+\frac{\eta}{2}\right)\sinh\left(\xi^--u-\frac{\eta}{2}\right)\right.\nonumber\\
&-&\left. (\kappa_-)^2\sinh^2(2u+\eta)\right\}\,. 
\eea
Here the parameters $\kappa_\pm$, $\xi_\pm$ are defined by the $K$-matrix \eqref{k_matrix_nondiag}. Note that the diagonal case \eqref{k_matrix_diag} is simply recovered by setting $\kappa_\pm \to 0$.

Note that since the transfer matrices $T^{(n)}(u)$ are commuting operators, they share a basis of common eigenvectors and the functional relation \eqref{t_system} holds also at the level of the corresponding eigenvalues. This is also the case for other relations written in this section involving the transfer matrices $T^{(n)}(u)$.

Next, from the $T$-system \eqref{t_system} one can derive a new set of functional relations which provides the well-known $Y$-system \cite{kp-92,kns-11}. Introducing the so-called $y$-functions
\be
y_j(u)=\frac{T^{(j-1)}(u)T^{(j+1)}(u)}{\Phi_j(u)}\,,
\label{y_function}
\ee
the $Y$-system reads
\be
y_j\left(u+\frac{\eta}{2}\right)y_j\left(u-\frac{\eta}{2}\right)=\left[1+y_{j+1}\left(u\right)\right]\left[1+y_{j-1}\left(u\right)\right]\,,
\label{y_system}
\ee
where $y_0\equiv 0$. 

As we already mentioned, the importance of this construction for our purposes lies in the possibility of casting the $Y$-system \eqref{y_system} into the form of a set of integral equations, which can then be solved numerically. In turn, this gives access to the set of $T$-functions $T^{(n)}(u)$, which include the boundary transfer matrix \eqref{QTMdef}. As we already stressed, the functional relation \eqref{t_system} holds also at the level of the eigenvalues of the transfer matrices $T^{(n)}(u)$. As a consequence, this opens the possibility of computing the leading eigenvalue of the transfer matrix \eqref{QTMdef}.

The derivation of a set of non-linear integral equations corresponding to the $Y$-system \eqref{y_system} is standard (see for example \cite{fgsw-11}), and we briefly review it here. It is first convenient to introduce the functions $\tilde{y}_j(\lambda)$ defined on the rotated plane $\lambda=iu$, namely
\be
\tilde{y}_j(\lambda)=y_j(-i\lambda)\,,
\label{eq:rotated_y}
\ee
which satisfy the $Y$-system
\be
\tilde{y}_j\left(\lambda+i\frac{\eta}{2}\right)\tilde{y}_j\left(\lambda-i\frac{\eta}{2}\right)=\left[1+\tilde{y}_{j+1}\left(\lambda\right)\right]\left[1+\tilde{y}_{j-1}\left(\lambda\right)\right]\,.
\label{y_system_lambda}
\ee
It is easy to see that the functions $\tilde{y}_j(\lambda)$ are periodic along the real direction with the corresponding period equal to $\pi$. It is also convenient to introduce the conventions employed in this work for the Fourier series expansion of a $\pi$-periodic function, which will be used shortly. They read
\bea
\hat{f}(k) &=& \int_{-\pi/2}^{\pi/2} \mathrm{d}\lambda e^{2 i k \lambda} f(\lambda) \,, \qquad k\in \mathbb{Z} \,, \label{fourier_1} \\
f(\lambda) &=& \frac{1}{\pi}\sum_{k \in \mathbb{Z}} e^{- 2 i k \lambda} \hat{f}(k) \,, \qquad \lambda \in \mathbb{R} \,. \label{fourier_2}
\eea

The precise form of the integral equations that we wish to derive depends on the analytical structure of the $y$-functions inside the so-called physical strip. The latter is the subset of the complex plane defined by
\be
\mathcal{S} = \left\{\lambda \Big| -\frac{\pi}{2} \leq \Re \lambda \leq \frac{\pi}{2} \,, -\frac{\eta}{2} \leq \Im \lambda \leq \frac{\eta}{2}\right\} \,,
\label{physicalstrip} 
\ee
where $\Re \lambda$ and $\Im\lambda$ denote respectively the real and imaginary part of the complex number $\lambda$. 

\begin{figure}
\centering
\begin{tikzpicture}
\begin{scope}
\fill[blue!10] (-2,-1) rectangle (2,1);

\fill[red] (1.3,0) circle[radius=0.05];
\fill[red] (2,.5) circle[radius=0.05];
\fill[red] (-2,.5) circle[radius=0.05];
\fill[red] (-0.65,1) circle[radius=0.05];
\fill[red] (-0.25, .5) circle[radius=0.05];
\fill[red] (1,-.6) circle[radius=0.05];

\draw[gray] (0,1) node[above, black] {\footnotesize $i \frac{\eta}{2}$}  -- (0,-1) node[below, black] {\footnotesize $-i  \frac{\eta}{2}$};
\draw[gray] (-2,0) node[left, black] {\footnotesize $-\frac{\pi}{2}$}   -- (2,0) node[right, black] {\footnotesize $\frac{\pi}{2}$} ;
\draw[line width=1,blue,->] (-2,1) -- (-1.5,1);
\draw[line width=1,blue] (-1.6,1) -- (2,1);
\end{scope}

\node at (3.1,0) {$=$};

\begin{scope}[shift={(5.5,0)}]
\fill[blue!10] (-2,-1) rectangle (2,1);
\fill[red] (1.3,0) circle[radius=0.05];
\fill[red] (2,.5) circle[radius=0.05];
\fill[red] (-2,.5) circle[radius=0.05];
\fill[red] (-0.65,1) circle[radius=0.05];
\fill[red] (-0.25, .5) circle[radius=0.05];
\fill[red] (1,-.6) circle[radius=0.05];

\draw[gray] (0,1)  -- (0,-1) ;
\draw[gray] (-2,0)  -- (2,0) ;
\draw[line width=1,blue,->] (-2,0) -- (-1.5,0);
\draw[line width=1,blue] (-1.6,0) -- (2,0);
\end{scope}

\node at (8,0) {$+$};

\begin{scope}[shift={(10.5,0)}]
\fill[blue!10] (-2,-1) rectangle (2,1);

\fill[red] (1.3,0) circle[radius=0.05];
\fill[red] (2,.5) circle[radius=0.05];
\fill[red] (-2,.5) circle[radius=0.05];
\fill[red] (-0.65,1) circle[radius=0.05];
\fill[red] (-0.25, .5) circle[radius=0.05];
\fill[red] (1,-.6) circle[radius=0.05];

\draw[gray] (0,1)  -- (0,-1) ;
\draw[gray] (-2,0)  -- (2,0) ;
\draw[line width=1,blue,->] (-2,1) -- (-1.5,1);
\draw[line width=1,blue] (-1.6,1)  -- (-0.8,1);
\draw[line width=1,blue] (-0.5,1)-- (2,1) --(2,0.65);
\draw[line width=1,blue] (2,0.35) -- (2,0) -- (1.45,0);
\draw[line width=1,blue]  (1.15,0) -- (-2,0) -- (-2,0.35);
\draw[line width=1,blue] (-2,0.65) --(-2,1);
\draw[line width=1,blue, rounded corners=3pt] (1.45,0) -- (1.3,0.15) -- (1.15,0) ;
\draw[line width=1,blue, rounded corners=3pt] (-0.8,1) -- (-0.65,0.85) -- (-0.5,1) ;
\draw[line width=1,blue, rounded corners=3pt] (-2,0.35) -- (-1.85,0.5) -- (-2,0.65) ;
\draw[line width=1,blue, rounded corners=3pt] (2,0.35) -- (1.85,0.5) -- (2,0.65) ;
\end{scope}

\node at (3.6,-3) {$+\frac{1}{2}$};
\begin{scope}[shift={(6,-3)}]
\fill[blue!10] (-2,-1) rectangle (2,1);
\draw[gray] (0,1)  -- (0,-1) ;
\draw[gray] (-2,0)  -- (2,0) ;
\draw[line width=1,blue, rounded corners=3pt]  (1.3,0) circle[radius=0.15];
\draw[line width=0.75,blue, rounded corners=3pt,->] (1.3,0.15) -- (1.35,0.15);

\draw[line width=1,blue, rounded corners=3pt]  (-0.65,1) circle[radius=0.15];
\draw[line width=0.75,blue, rounded corners=3pt,->] (-0.65,0.85) -- (-0.7,0.85);

\draw[line width=1,blue, rounded corners=3pt]  (2,0.5) circle[radius=0.15];
\draw[line width=0.75,blue, rounded corners=3pt,->] (1.85,0.5) -- (1.85,0.55);
\draw[line width=1,blue, rounded corners=3pt]  (-2,0.5) circle[radius=0.15];
\draw[line width=0.75,blue, rounded corners=3pt,-<] (-1.85,0.5) -- (-1.85,0.55);

\fill[red] (1.3,0) circle[radius=0.05];
\fill[red] (2,.5) circle[radius=0.05];
\fill[red] (-2,.5) circle[radius=0.05];
\fill[red] (-0.65,1) circle[radius=0.05];
\fill[red] (-0.25, .5) circle[radius=0.05];
\fill[red] (1,-.6) circle[radius=0.05];
\end{scope}
\end{tikzpicture}
\caption{
Procedure used to move the integration paths. Here we show for instance how to go from the integration path $\int_{- \frac{\pi}{2} + i \frac{\eta}{2}}^{\frac{\pi}{2} + i \frac{\eta}{2}}  \mathrm{d}\lambda $ to the path $\int_{- \frac{\pi}{2}}^{\frac{\pi}{2}}  \mathrm{d}\lambda $, while picking some residues at the poles of the integrand (represented as red dots). An analogous procedure is taken for $\int_{- \frac{\pi}{2} - i \frac{\eta}{2}}^{\frac{\pi}{2} - i \frac{\eta}{2}}  \mathrm{d}\lambda $.
The shaded area is the physical strip \eqref{physicalstrip}.  
}
 \label{fig:integrationcontours}
\end{figure}
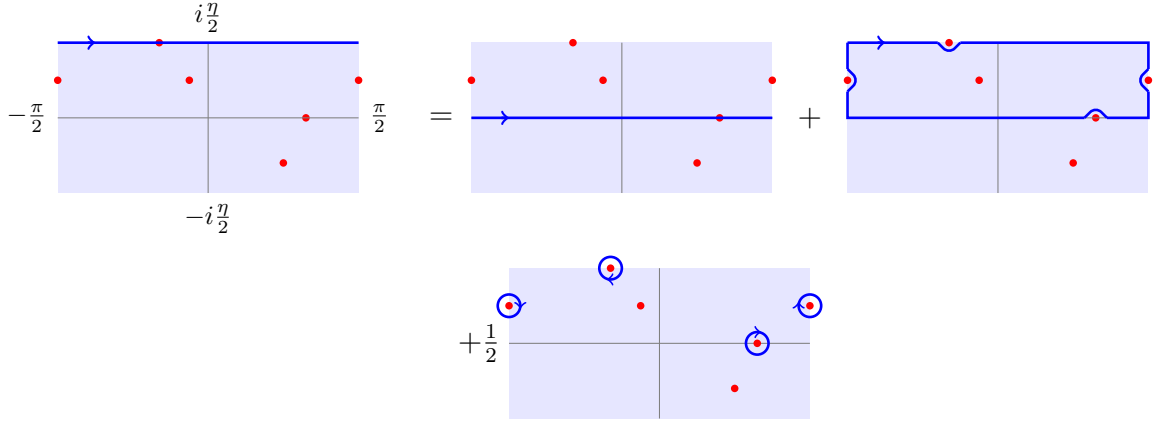

The steps necessary to cast the $Y$-system \eqref{y_system_lambda} into the form of non-linear integral equations can then be summarized as follows. First, we take the logarithmic derivative on both sides of \eqref{y_system_lambda} and Fourier transform them. The integral appearing in the l. h. s. are now along segments with non-zero imaginary parts. These can be moved in the physical strip back to the real line, taking care of the poles of the logarithmic derivates as pictorially represented in Fig.~\ref{fig:integrationcontours}. One is therefore left with equations of the form 
\be
\reallywidehat{\log \tilde{y}_j} = \frac{1}{2\cosh k \eta}  \left[
\reallywidehat{\log\left(1+\tilde{y}_{j+1}\right)} 
+
\reallywidehat{\log\left(1+\tilde{y}_{j-1}\right)}  
 \right]
 + 
 \ldots \,, 
\ee
where the $\ldots$ denote additional contributions coming from the poles. Such equations can then be Fourier transformed back to real space, yielding the desired set of non-linear integral equations.

We are thus left with the problem of understanding the analytical structure of the rotated $y$-functions \eqref{eq:rotated_y} inside the physical strip \eqref{physicalstrip}. This analysis has to be performed separately for each state of interest. We perform this explicitly in the following for the cases of the N\'eel state and the tilted N\'eel state, which respectively provide an example of diagonal and non-diagonal reflection matrices. For these states we explicitly derive the corresponding set of non-linear integral equations. These will be then explicitly solved in section~\ref{sec:resolution} where numerical results for the dynamical free energy and Loschmidt echo will be presented.

\subsection{The N\'eel state}

We now study the $Y$-system associated to the leading eigenvalue of the transfer matrix \eqref{QTMdef} in the case of the N\'eel state.
Equations (\ref{y_function}), (\ref{t_system}), and \eqref{eq:first:t} immediately yield the first $y$-function
\be
1+y_1(u)=\frac{T_{\bm{\lambda}}\left(u+\eta/2\right)T_{\bm{\lambda}}\left(u-\eta/2\right)}{f(u-\eta/2)}\,,
\label{explicit_t}
\ee
where we have indicated with $T_{\bm{\lambda}}(u)$ the leading eigenvalue of the transfer matrix \eqref{QTMdef} associated with rapidities $\{\lambda\}_{j=1}^N$. Here we explicitly used that the $T$-system \eqref{t_system} holds separately for each common eigenstate of the fused transfer matrices.

In \eqref{explicit_t}, the function $f(u)$ is defined in \eqref{f_function}. In the present case one has $\kappa_\pm = 0$, $\xi_\pm = \mp \eta/2$, so
\be
f(u - \eta/2) =  \phi(u+\eta) \phi(u-\eta) \omega_1 (u+\eta/2) \omega_2 (u-\eta/2)  \,,
\label{fdiagonal}
\ee
where $\phi(u)$, $\omega_1(u)$ and $\omega_2(u)$ are defined in (\ref{phi_function}), (\ref{omega1_function}), (\ref{omega2_function}). Using (\ref{TQrelation_diagonal}) and (\ref{explicit_t}) we immediately get the following expression of $y_1$
\be
y_1(u)=\fa(u+\eta/2)+\frac{1}{\fa(u-\eta/2)}+\frac{\fa(u+\eta/2)}{\fa(u-\eta/2)}\,,
\label{eq:y_1_diag}
\ee
in terms of the auxilliary function
\bea
\hspace{-2cm}
\fa(u)
&=&
\frac{\omega_2(u)}{\omega_1(u)}\frac{Q(u+\eta)\phi(u-\eta/2)}{Q(u-\eta)\phi(u+\eta/2)}\,.
\nonumber \\
\hspace{-2cm}
&=&
K(u)
\left[
\frac{\sinh(u+\beta/2N-\eta)}{\sinh(u-\beta/2N+\eta)}
\frac{\sinh(u-\beta/2N)}{\sinh(u+\beta/2N)}
\right]^{2N}
\prod_{k=1}^{2N} \frac{\sinh(u-\tilde \lambda_k+\eta)}{\sinh(u-\tilde \lambda_k-\eta)}\,,
\label{a_q} 
\eea
with
\be
K(u)=\frac{\sinh(u+\eta)\sinh(2u-\eta)}{\sinh(u-\eta)\sinh(2u+\eta)}\,.
\label{k_function}
\ee

Equation \eqref{eq:y_1_diag} gives the exact expression for $y_1(u)$ at finite Trotter number $N$, and hence of the rotated function $\tilde{y}_1(\lambda)$ in \eqref{eq:rotated_y}. We have studied numerically the analytical structure of the latter for Trotter number $2N= 2,4,6,8$. Based on this, and analytical inspection, we conjecture the validity of the following analytical structure of poles and zeros inside the physical strip for general $N$. 
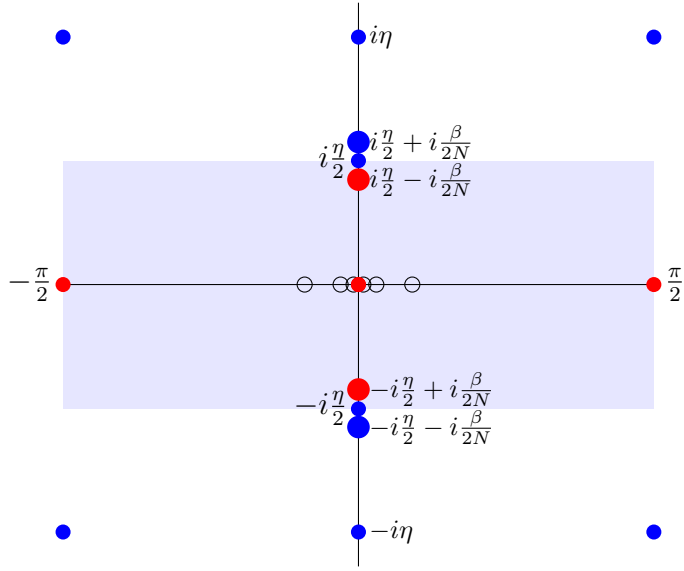
\begin{figure}
\centering
\begin{tikzpicture}
\begin{scope}[scale=2.5]

\fill[blue!10] (-1.5708,-0.658479) rectangle (1.5708,0.658479);

\draw (-1.5708,0) -- (1.5708,0);
\draw (0,-1.5) -- (0,1.5);
 
\foreach \x in {(-0.286183,0),(0.286183,0),(-0.0949507,0),(0.0949507,0),(-0.0260309,0),(0.0260309,0)}
  \draw \x circle[radius=0.04];

\foreach \x in {(-1.5708,0),(1.5708,0),(0,0),(0,0)}
  \fill[red] \x circle[radius=0.04];
 \fill[red]  (0,-0.558479) circle[radius=0.06];
 \fill[red]  (0,0.558479) circle[radius=0.06];

\foreach \x in {(-1.5708,1.31696),(-1.5708,-1.31696),(1.5708,1.31696),(1.5708,-1.31696),(0,-1.31696),(0,-0.661555),(0,0.659383),(0,1.31696)
}
  \fill[blue] \x circle[radius=0.04];
  
   \fill[blue]  (0,-0.758479) circle[radius=0.06];
 \fill[blue]  (0,0.758479) circle[radius=0.06];

\node[right] at (1.5708,0) {$\frac{\pi}{2}$};
\node[left] at (-1.5708,0) {$-\frac{\pi}{2}$};

\node[left] at (0,-0.658479) {$-i\frac{\eta}{2}$};
\node[right] at (0.,-0.758479) {\footnotesize $-i\frac{\eta}{2}-i\frac{\beta}{2N}$};
\node[right] at (0.,-0.558479) {\footnotesize $-i\frac{\eta}{2}+i\frac{\beta}{2N}$};

\node[left] at (0,0.658479) {$i\frac{\eta}{2}$};
\node[right] at (0.,0.758479) {\footnotesize $i\frac{\eta}{2}+i\frac{\beta}{2N}$};
\node[right] at (0.,0.558479) {\footnotesize $i\frac{\eta}{2}-i\frac{\beta}{2N}$};

\node[right] at (0.,1.31696) {\footnotesize $i\eta$};
\node[right] at (0.,-1.31696) {\footnotesize $-i\eta$};
\end{scope}

\end{tikzpicture}
\caption{Poles (blue dots) and zeroes (red dots) of the rotated $y$-function $\tilde{y}_1(\lambda)$ associated to the largest quantum transfer matrix eigenvalue with diagonal boundary conditions. The parameters are chosen as $2N=6$, $\Delta=2$, $\beta=0.6$, $\xi_\pm = \mp \eta/2$. 
The shaded area is the physical strip (\ref{physicalstrip}), while multiple poles and zeroes are represented as larger dots. The corresponding (doubled) Bethe roots $\tilde{\lambda}_j$ are also displayed for completeness (empty circles). Note that in the rotated complex plane they lie on the real axis. Some additional zeroes and poles not represented here exist further away from the physical strip, such as poles at $\lambda=i\left(\mp 3\eta/2\pm \beta/2N\right)$.
}
 \label{fig:polesdiag}
\end{figure}
We consider for simplicity the case $\beta\geq 0$, while a similar analysis holds for $\beta<0$. The general structure is the following:
\begin{itemize}
\item $\tilde{y}_1(\lambda)$ has zeroes of order 2 independent of $N$ in $\lambda=0$, $\lambda=\pm \frac{\pi}{2}$ .
\item It has $N$-dependent zeroes of order $2N$ in $\lambda = i\left(\pm\eta/2\mp\beta/2N \right)$.
\item It has additional pairs of poles in $\pm i \eta/2$ .
\item It has no further zeroes or poles inside the physical strip.
\end{itemize}
Note also that
\begin{itemize}
\item{It has $N$-dependent poles in $\lambda= i\left( \pm\eta/2\pm\beta/2N \right)$, $\lambda =i \left(\mp 3\eta/2\pm \beta/2N \right)$. For $N$ large enough and $\beta \geq 0$ these however lie outside the physical strip, and do not contribute to the NLIE.}
\item{There are no zeroes or poles coinciding with the Bethe roots.}
\end{itemize}
This is illustrated in Fig.~\ref{fig:polesdiag} for Trotter number $2N=6$.
Based on this analytical structure, and following the prescription explained in the previous subsection, we can cast the functional relation
 \be
\tilde{y}_1(\lambda+i\eta/2) \tilde{y}_1(\lambda-i\eta/2) = 1 + \tilde{y}_2(\lambda) \,,
\ee
into integral form. Note that the contribution to the integral equation of pairs of poles or zeros located at $ \lambda = \pm i \eta/2$ cancel. The final result reads

\be
\log[\tilde{y}_1(\lambda)]=\varepsilon[\beta, N](\lambda)+d_1(\lambda)+\left[s\ast\log\left(1+\tilde{y}_2\right)\right](\lambda)\,,
\label{first_NLIE_diagonal}
\ee
where we used the definitions
\bea
\label{dn-functions}
d_n(\lambda)&=&\sum_{k\in \mathbb{Z}}e^{-2ik\lambda}\frac{\tanh(k\eta)}{k}\left[(-1)^n-(-1)^k\right]\,, \label{d_n_function}\\
\varepsilon[\beta,N](\lambda)&=&
-
2N\sum_{k\in \mathbb{Z}}e^{-2ik\lambda}\frac{\sinh(k\beta/N)}{k\cosh{(k\eta)}}\,,\label{epsilon_function}
\eea
and
\be
s(\lambda)=\frac{1}{2\pi}\sum_{k\in\mathbb{Z}}\frac{e^{-2ik\lambda}}{\cosh(k\eta)}\,.
\label{s_function}
\ee
We also introduced the following notation for the convolution of two functions
\be
\left[g\ast h\right] (\lambda)=\int_{-\pi/2}^{+\pi/2}\,{\rm d}\mu\  g(\lambda-\mu)h(\mu)\,.
\label{convolution}
\ee
We note that the functions $d_n(\lambda)$ can be written using the
Jacobi-theta functions as \cite{wdbf-14_2}
\be
\label{jacobi}
d_n(\lambda)=(-1)^n \log\frac{\vartheta^2_4(\lambda)}{\vartheta^2_1(\lambda)}
+\log\frac{\vartheta^2_2(\lambda)}{\vartheta^2_3(\lambda)}\,,
\ee
where the nome is $e^{-2\eta}$.

It is useful to compare the analytic structure to that of the purely
thermal case \cite{klumper-04}. The new additions are the $N$-independent zeroes of
$\tilde y_1$ resulting in the extra source term $d_1(\lambda)$, which
is not present in the usual TBA. Note also that in principle a non-vanishing constant of integration might be present in the r. h. s. of \eqref{first_NLIE_diagonal}. However, in analogy with the thermal case, we can convince ourselves that such constant is zero. This is also a posteriori checked by the excellent agreement of our numerical solution of the TBA equations with exact diagonalization calculations.

We now turn to the integral equations corresponding to higher index $n$. From the $Y$-system \eqref{y_system_lambda}, at any finite Trotter number higher $y$-functions can be simply obtained from the knowledge of $\tilde{y}_1(\lambda)$. Analogously to the case $n=1$ we have then studied numerically the analytical structure inside the physical strip of $\tilde{y}_n$ for $n=2,3,4$ at Trotter number $2N=2,4,6,8$. A pattern clearly emerges, which is summarized as follows (restricting again for simplicity to $\beta\geq 0$)
\begin{itemize}
\item The $N$-dependent zeros $\lambda=i\left(\pm\eta/2\mp\beta/2N \right)$ are absent for all $n>1$. Furthermore, there are no additional $N$-dependent zeroes or poles for $n>1$. 
\item At $\lambda=0$, $\tilde{y}_n(\lambda)$ has a zero of order $2$ for $n$ odd, and a pole of order $2$ for $n$ even. 
\item At $\lambda=\pm \pi/2$, $\tilde{y}_n(\lambda)$ has a zero of order $2$.
\item There are additional poles at $\lambda=\pm i\eta/2$ which however do not contribute to the derivation of the integral equations.
\end{itemize}
Based on this analytical structure and following the prescription of the last section, the functional relation $\eqref{y_system_lambda}$ for $n>1$ is cast in the form
\bea
\log[\tilde{y}_n(\lambda)]&=&d_n(\lambda)+\left[s\ast\left\{\log\left(1+\tilde{y}_{n-1}\right)+\log\left(1+\tilde{y}_{n+1}\right)\right\}\right](\lambda)\,,
\label{finite_N_NLIE_diagonal}
\eea
where $d_n(\lambda)$ is defined in \eqref{d_n_function}.

It is now straightforward to compute the Trotter limit of equations \eqref{first_NLIE_diagonal}, \eqref{finite_N_NLIE_diagonal}, by noticing that
\be
\lim_{N\to\infty}\varepsilon[\beta,N](\lambda)=-4\pi\beta s(\lambda)\,,
\label{limit_s}
\ee
where $s(\lambda)$ is given in \eqref{s_function}. We thus arrive at the final result
\bea
\log[\tilde{y}_1(\lambda)]&=&-4\pi\beta s(\lambda)+d_1(\lambda)+\left[s\ast\log\left(1+\tilde{y}_2\right)\right](\lambda)\,,\nonumber\\
\log[\tilde{y}_n(\lambda)]&=&d_n(\lambda)+\left[s\ast\left\{\log\left(1+\tilde{y}_{n-1}\right)+\log\left(1+\tilde{y}_{n+1}\right)\right\}\right](\lambda)\,.
\label{TBA_diagonal}
\eea
These equations have been obtained for $\beta\geq 0$ but a similar derivation in the case $\beta<0$ shows that they hold for $\beta\in\mathbb{R}$.

We note that these have the typical form of the partially decoupled integral equations appearing in the thermodynamic Bethe ansatz (TBA) analysis of the $XXZ$ Heisenberg chain at thermal equilibrium \cite{takahashi-99}. 
Remarkably, setting $\beta=0$ we see that \eqref{TBA_diagonal} coincide with the so-called generalized TBA equations derived in \cite{wdbf-14,wdbf-14_2,pmwk-14} for the steady state in quenches from the N\'eel state. We will return to this in section~\ref{sec:qam} where we explicitly discuss the connection between the quantum transfer matrix formalism and the quench action approach employed in \cite{wdbf-14,wdbf-14_2,pmwk-14}.

In order to be solved, these equations have to be supplemented with an asymptotic condition for the behavior of $\tilde{y}_n(\lambda)$ at large $n$, which will be discussed in section~\ref{sec:resolution}.

\subsection{The tilted N\'eel state}

The derivation of non-linear integral equations in the case of non-diagonal boundary conditions follows closely the diagonal case previously discussed. Our starting point is once again given by equation \eqref{explicit_t}, which provides the first $y$-function associated with the leading eigenvalue $T_{\bm{\lambda}}(u)$ of the transfer matrix \eqref{QTMdef} with non-diagonal $K$-matrices \eqref{k_matrix_nondiag}. The latter is in this case associated with a set of $2N$ Bethe roots $\{\lambda_j\}_{j=1}^{2N}$.

The function $f(u)$ is now given in equation \eqref{f_function}. After straightforward calculations, one can show that $f(u)$ can be written as  
\bea
\hspace{-2cm} f(u-\eta/2) &=&
 (4 \kappa_+ \kappa_-)^2 \cosh(u-\beta_+)\cosh(u+\beta_+)\cosh(u-\beta_-)\cosh(u+\beta_-) \nonumber \\
\hspace{-3cm} &\times &  
\sinh(u+\alpha_+)\sinh(u-\alpha_+)\sinh(u+\alpha_-)\sinh(u-\alpha_-)
 \nonumber \\
\hspace{-2cm} &\times &
\frac{\sinh(2u+2\eta)\sinh(2u-2\eta)}{\sinh(2u+\eta)\sinh(2u-\eta)} 
\phi(u+\eta) \phi(u-\eta) \nonumber\\
&=&(\sinh\xi_+\sinh\xi_-)^2\mathbf{A}\left(u+\eta/2\right)\mathbf{A}\left(-u+\eta/2\right)\,,
\hspace{-2cm}
 \label{fnondiag}
\eea
where $\phi(u)$ and $\mathbf{A}(u)$ are defined in \eqref{phi_function} and \eqref{eq:bold_a_function} respectively. Note that in the diagonal limit $\beta_\pm = \zeta  \to \infty$, $\kappa_\pm \sim e^{-\zeta} \to 0$, $\alpha_\pm \to \xi_\pm$. Accordingly, it is easy to see that in this limit $f(u-\eta/2)$ coincides with \eqref{fdiagonal} as it should. 

Using the explicit expression \eqref{TQrelation_nondiagonal} for the leading eigenvalue $T_{\bm{\lambda}}(u)$ of the quantum transfer matrix, we can rewrite $y_1(u)$ as
\be
1+y_1(u)=1+\mathcal{F}_{\rm hom}(u)+\mathcal{F}_{\rm mix}(u)+\mathcal{F}_{\rm inhom}(u)\,.
\label{general_form}
\ee
The first term is the one generated from the homogeneous AQ-terms and has the form
\be
\mathcal{F}_{\rm hom}(u)=\fb(u+\eta/2)+\frac{1}{\fb(u-\eta/2)}+\frac{\fb(u+\eta/2)}{\fb(u-\eta/2)}\,,
\ee
where now we defined
\bea
\fb(u)=\frac{\mathbf{A}(-u)Q(u+\eta)}{\mathbf{A}(u)Q(u-\eta)}\,.
\label{aux_a}
\eea
Analogously, the third term in (\ref{general_form}) is the one generated by the product of the two inhomogenous $F$-terms. It reads
\bea
\mathcal{F}_{\rm inhom}(u)&=&\frac{1}{\mathbf{A}\left(u+\eta/2\right)\mathbf{A}\left(-u+\eta/2\right)}\frac{F(u-\eta/2)F(u+\eta/2)}{Q(u-\eta/2)Q(u+\eta/2)}\,.
\eea
Finally, $\mathcal{F}_{\rm mix}(u)$ is given by the product of the homogeneous and inhomogeneous terms and is defined as
\bea
\mathcal{F}_{\rm mix}(u)&=&\frac{F(u-\eta/2)}{\mathbf{A}\left(-u+\eta/2\right)}\frac{1}{Q(u+\eta/2)}\left(1+\fb(u+\eta/2)\right)\nonumber\\
&+& \frac{F(u+\eta/2)}{\mathbf{A}\left(u+\eta/2\right)}\frac{1}{Q(u-\eta/2)}\left(1+\frac{1}{\fb(u-\eta/2)}\right)
\,.
\eea

Analogously to the diagonal case, \eqref{general_form} gives the exact
expression for $y_1(u)$ at finite Trotter number $N$, and hence of the
rotated function $\tilde{y}_1(\lambda)$. We studied the analytical structure of
the latter for Trotter number $2N=2,4,6$, from which
a clear pattern has emerged, allowing us to formulate a conjecture one
the analytical structure of poles and zeros inside the physical strip
for general $N$.  
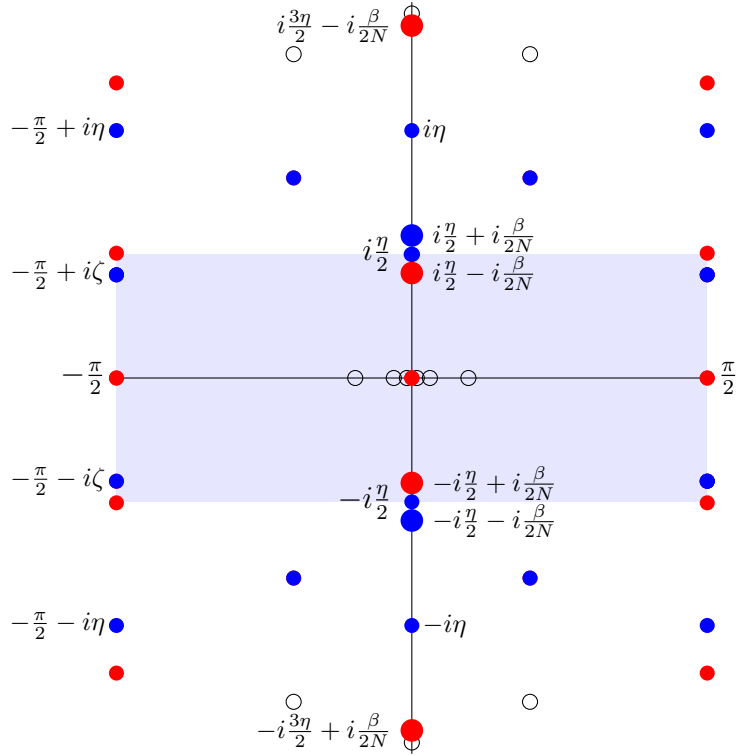
\begin{figure}
\centering
\begin{tikzpicture}
\begin{scope}[scale=2.5]

\fill[blue!10] (-1.5708,-0.658479) rectangle (1.5708,0.658479);

\draw (-1.5708,0) -- (1.5708,0);
\draw (0,-2) -- (0,2);
 
\foreach \x in {(-0.628511, -1.72327), (0.628511, -1.72327), (0, -1.93988),(-0.30092, 0), (0.30092, 0), (-0.0961014, 0), (0.0961014,0), (-0.0262106, 0), (0.0262106, 0), (-0.628511,1.72327), (0.628511, 1.72327), (0, 1.93988)}
  \draw \x circle[radius=0.04];

  
    \foreach \x in {
(1.5708, 1.57034), (1.5708, 0.664216), (1.5708, 0.549306), (1.5708, 0),(1.5708, 0),(1.5708, -0.549306), (1.5708, -0.664216), (1.5708, -1.57034),(-1.5708, 1.57034), (-1.5708, 0.664216),(-1.5708, 0.549306), (-1.5708, 0),(-1.5708, 0), (-1.5708, -0.549306), (-1.5708, -0.664216), (-1.5708, -1.57034),  (-0.628511, -1.06479), (0.628511, -1.06479),  
  (-0.0000612281, 0), (0.0000612281,   0), (-0.628511, 1.06479), (0.628511,   1.06479),
     }
      \fill[red] \x circle[radius=0.04];

   \fill[red]  (0,-0.558479) circle[radius=0.06];
 \fill[red]  (0,0.558479) circle[radius=0.06];
  \fill[red]  (0,-1.87544) circle[radius=0.06];
   \fill[red]  (0,1.87544) circle[radius=0.06];

\foreach \x in {
(1.5708, 1.31696),(-1.5708, 1.31696), (-1.57077, 0.549327), (1.57077, 0.549327),(-1.57077, 0.549285), (1.57077, 0.549285), (1.5708, -0.549277),(-1.57077, -0.549306), (1.57077, -0.549306), (1.5708, -0.549335),(1.5708, -1.31696),(-1.5708, -1.31696), (-0.628511, -1.06479), (0.628511,-1.06479), (0, -1.31696),  (0,-0.659237), (-0.628511, 1.06479), (0.628511, 1.06479),(-0.00354966, 0.658125), (0.00354966, 0.658125), (0, 1.31696)
}
  \fill[blue] \x circle[radius=0.04];

   \fill[blue]  (0,-0.758479) circle[radius=0.06];
 \fill[blue]  (0,0.758479) circle[radius=0.06];

\node[right] at (1.5708,0) {$\frac{\pi}{2}$};
\node[left] at (-1.5708,0) {$-\frac{\pi}{2}$};

\node[left] at (-0.05,-0.658479) {$-i\frac{\eta}{2}$};
\node[right] at (0.05,-0.758479) {\footnotesize $-i\frac{\eta}{2}-i\frac{\beta}{2N}$};
\node[right] at (0.05,-0.558479) {\footnotesize $-i\frac{\eta}{2}+i\frac{\beta}{2N}$};

\node[left] at (-0.05,0.658479) {$i\frac{\eta}{2}$};
\node[right] at (0.05,0.758479) {\footnotesize $i\frac{\eta}{2}+i\frac{\beta}{2N}$};
\node[right] at (0.05,0.558479) {\footnotesize $i\frac{\eta}{2}-i\frac{\beta}{2N}$};

\node[left] at (-0.05,1.87544) {\footnotesize $i\frac{3\eta}{2}-i\frac{\beta}{2N}$};
\node[left] at (-0.05,-1.87544) {\footnotesize $-i\frac{3\eta}{2}+i\frac{\beta}{2N}$};

\node[right] at (0.,1.31696) {\footnotesize $i\eta$};
\node[right] at (0.,-1.31696) {\footnotesize $-i\eta$};

\node[left] at (-1.57077, 0.549285) {\footnotesize $-\frac{\pi}{2}+ i \zeta$};
\node[left] at (-1.57077, -0.549285) {\footnotesize $-\frac{\pi}{2}- i \zeta$};

\node[left] at (-1.57077,1.31696) {\footnotesize $-\frac{\pi}{2}+ i \eta$};
\node[left] at (-1.57077,-1.31696) {\footnotesize $-\frac{\pi}{2}-  i \eta$};

\end{scope}

\end{tikzpicture}
\caption{Poles (blue dots) and zeroes (red dots) of the rotated function $\tilde{y}_1(\lambda)$ associated to the largest quantum transfer matrix eigenvalue with non-diagonal boundary conditions corresponding to the tilted N\'eel with $\vartheta = \frac{\pi}{3}$. The parameters are chosen as $2N=6$, $\Delta=2$, $\beta=0.6$, $\xi_\pm = \pm \eta/2$, $\zeta \simeq 0.55$. The shaded area is the physical strip (\ref{physicalstrip}), while multiple poles and zeroes are represented as larger dots. The corresponding (doubled) Bethe roots $\tilde{\lambda}_j$ are also displayed for completeness (empty circles).  Note that in the rotated complex plane the rapidities $\tilde{\lambda}_j^{\rm reg}$ are displaced on the real axis.}
 \label{fig:polesnondiag}
\end{figure}

In summary, the analytical structure of the rotated function $\tilde{y}_1(\lambda)$ in the general case is very similar to that described in the previous section for the diagonal case, cf. Fig.~\ref{fig:polesnondiag}. In particular, $\tilde{y}_1(\lambda)$ displays the same poles and zeroes as the corresponding function in the diagonal case discussed in the previous section. In addition, in the non-diagonal case $\tilde{y}_1(\lambda)$ has also poles of order $2$ at 
\be
\lambda  = \frac{\pi}{2} \pm i \zeta \,,\qquad \lambda  = -\frac{\pi}{2} \pm i \zeta\,.
\ee
Whether these poles lie within the physical strip (\ref{physicalstrip}) and therefore contribute to the first integral equation, depends on the values of $\zeta$ and $\eta$. 

Numerical examination for $2N=2,4,6$ of the analytical structure of $\tilde{y}_{n}(\lambda)$ for higher values of $n$ shows that one has a similar picture also in these cases. More precisely, our analysis has led us to the following conjecture for the analytical structure of the $y$-functions $\tilde{y}_n(\lambda)$ for the tilted N\'eel state.

\begin{itemize}
\item{The functions $\tilde{y}_n(\lambda)$ of the tilted N\'eel state have all the poles and zeroes corresponding to the $y$-functions of the N\'eel state discussed in the previous subsection.}
\item{In addition to these poles and zeroes, the functions $\tilde{y}_n(\lambda)$ of the tilted N\'eel state might also display new poles of order $2$ inside the physical strip, depending on the tilting angle $\vartheta$ and hence on the parameter $\zeta$ introduced in \eqref{def_zeta}.}
\end{itemize}
For each $\tilde{y}_n(\lambda)$ we now summarize the position of these extra poles.
We distinguish two cases.

\paragraph{Case 1 : $p \eta \leq \zeta < \left( p + \frac{1}{2}\right)\eta$ ~($p \in \mathbb{N}$)}

\begin{itemize}
\item For $n \leq 2p$, $\tilde{y}_n(\lambda)$ does not have extra poles or zeros in the physical strip.
\item For $n$ odd $\geq 2p+1$, $\tilde{y}_n$ has poles or order 2 at 
\be
\lambda = \frac{\pi}{2} \pm i \left(\zeta - p \eta \right)\,,\quad \lambda = -\frac{\pi}{2} \pm i \left(\zeta - p \eta \right)
\ee
\item for $n$ even $\geq 2p+2$, $\tilde{y}_n(\lambda)$ has poles or order 2 at 
\be
\fl \lambda = \frac{\pi}{2} \mp i \left[\zeta - \left(p + \frac{1}{2}\right) \eta \right] \,,\quad \lambda = -\frac{\pi}{2} \mp i \left[\zeta - \left(p + \frac{1}{2}\right) \eta \right]\,.
\ee
\end{itemize} 

\paragraph{Case 2 : $\left( p + \frac{1}{2}\right) \eta \leq \zeta < \left( p +1\right)\eta$ ~($p \in \mathbb{N}$)}

\begin{itemize}
\item For $n \leq 2p+1$, $\tilde{y}_n$ does not have extra poles in the physical strip.
\item For $n$ odd $\geq 2p+3$, $\tilde{y}_n$ has poles or order 2 at 
\be
\lambda = \frac{\pi}{2} \mp i \left[\zeta - (p+1) \eta \right]\,,\quad \lambda = - \frac{\pi}{2} \mp i \left[\zeta - (p+1) \eta \right]\,.
\ee
\item for $n$ even $\geq 2p+2$, $\tilde{y}_n$ has poles or order 2 at 
\be
\fl \lambda = \frac{\pi}{2} \pm i \left[\zeta - \left(p + \frac{1}{2}\right) \eta \right] \,,\qquad \lambda = -\frac{\pi}{2} \pm i \left[\zeta - \left(p + \frac{1}{2}\right) \eta \right]\,.
\ee
\end{itemize}

It is now straightforward to follow the prescription of the previous subsection to cast the functional relation \eqref{y_system_lambda} into the form of integral equations. We obtain
\bea
\fl \log[\tilde{y}_1(\lambda)]&=&\varepsilon[\beta, N](\lambda)+d_1(\lambda)+
\delta_{1}(\lambda)
+ 
\left[s\ast\log\left(1+\tilde{y}_2\right)\right](\lambda)\,,\\
\fl \log[\tilde{y}_n(\lambda)]&=&
d_n(\lambda) 
+
\delta_{n}(\lambda)+\left[s\ast\left\{\log\left(1+\tilde{y}_{n-1}\right)+\log\left(1+\tilde{y}_{n+1}\right)\right\}\right](\lambda)\,,
\eea
where $d_n(\lambda)$, $\varepsilon[\beta,N]$, $s(\lambda)$ are defined in \eqref{d_n_function}, \eqref{epsilon_function} and \eqref{s_function}. The driving terms  $\delta_{n}(\lambda)$ are generated by the additional poles described above. Accordingly, their definition depends on the value of $\zeta$ in \eqref{def_zeta}. As before, we distinguish two cases: 
\paragraph{Case 1 : $p \eta \leq \zeta < \left( p + \frac{1}{2}\right)\eta$ ~($p \in \mathbb{N}$).} In thise case, the following definitions hold
\bea
\fl \delta_{n}(\lambda)  =  0 \,,  \quad(n \leq 2p)  \nonumber \\  
\fl \delta_{n}(\lambda)  =   - 2\sum_{k\in \mathbb{Z}}e^{-2ik\lambda} \frac{(-1)^k}{k \cosh(k \eta)} \sinh\left( k( 2\zeta - (2p+1)\eta) \right)  \,, \quad(n~{\rm odd}~ \geq 2p+1) \nonumber \\ 
\fl \delta_{n}(\lambda) =  - 2\sum_{k\in \mathbb{Z}}e^{-2ik\lambda} \frac{(-1)^k}{k \cosh(k \eta)} \sinh\left( k(-2\zeta +2p \eta) \right)  \,, \quad(n~{\rm even}~ \geq 2p+2) \,.
\eea 

\paragraph{Case 2 :  $\left( p + \frac{1}{2}\right) \eta \leq \zeta < \left( p +1\right)\eta$ ~($p \in \mathbb{N}$).} In thise case, we have instead
\bea
\fl \delta_{n}(\lambda)  =   0   \quad(n \leq 2p+1)  \nonumber \\ 
\fl \delta_{n}(\lambda)  =   - 2\sum_{k\in \mathbb{Z}}e^{-2ik\lambda} \frac{(-1)^k}{k \cosh(k \eta)} \sinh\left( k(- 2\zeta + (2p+1)\eta) \right)   \quad(n~{\rm odd}~ \geq 2p+3) \nonumber \\ 
\fl \delta_{n}(\lambda)  =  - 2\sum_{k\in \mathbb{Z}}e^{-2ik\lambda} \frac{(-1)^k}{k \cosh(k \eta)} \sinh\left( k(2\zeta -(2p+2) \eta) \right)   \quad(n~{\rm even}~ \geq 2p+2)\,.
\eea 

Note that the Trotter limit can now be again computed straightforwardly by means of \eqref{limit_s}. We then arrive at the final result
\bea
\fl \log[\tilde{y}_1(\lambda)]&=&-4\pi\beta s(\lambda)+d_1(\lambda)+
\delta_{1}(\lambda)
+ 
\left[s\ast\log\left(1+\tilde{y}_2\right)\right](\lambda)\,,\nonumber\\
\fl \log[\tilde{y}_n(\lambda)]&=&
d_n(\lambda) 
+
\delta_{n}(\lambda)+\left[s\ast\left\{\log\left(1+\tilde{y}_{n-1}\right)+\log\left(1+\tilde{y}_{n+1}\right)\right\}\right](\lambda)\,.
\label{TBA_nondiagonal}
\eea

Note again that in principle a non-vanishing constant of integration might be present in the r. h. s. of these equations. As in the diagonal case, we can convince ourselves that it is zero, in analogy with the thermal case. These equations generalize \eqref{TBA_diagonal} to the case of the tilted N\'eel state. Again, they have to be supplemented with an asymptotic condition for $\tilde{y}_n(\lambda)$ at large $n$, which will be discussed in section~\ref{sec:resolution}. Instead, we now show how the solution of these infinite sets of integral equations gives immediately the leading eigenvalue of the transfer matrix \eqref{eq:mathcal_t} in the Trotter limit $N\to\infty$ and hence the dynamical free energy \eqref{eq:g_en_function}.

\subsection{From the $Y$-system to the dynamical free energy}

We now show how the leading eigenvalue of the transfer matrix \eqref{BQTMdef} can be directly obtained from the solution of the TBA-like integral equations derived in the last subsections.

We start by defining a function of a complex parameter $\lambda$ which coincides with $\mathcal{T}$ in \eqref{eq:mathcal_t} for $\lambda=0$. With a slight abuse of notations, we call such a function $\mathcal{T}(\lambda)$ and define it as 
\bea 
 \mathcal{T}(\lambda)=\frac{1}{\left[\sin\left(\lambda-i\frac{\beta}{2N}+i\eta\right)\sin\left(-\lambda-i\frac{\beta}{2N}+i\eta\right)\right]^{2N}}\frac{1}{\mathcal{N}(\lambda)}
T(i\lambda)
\,, 
\label{BQTMlambdadef}
\eea 
where
\be
\mathcal{N}(\lambda)=\langle v^+(i\lambda)|v^-(i\lambda)\rangle\,,
\label{eq:norm}
\ee
and where $|v^\pm(u)\rangle$ are defined in \eqref{v1}, \eqref{v2}. Analogously, we denote with $\Lambda_0(\lambda)$ the eigenvalue of $\mathcal{T}(\lambda)$ such that $\Lambda_0(0)=\Lambda_0$ is consistently the leading eigenvalue of $\mathcal{T}(0)$ in \eqref{BQTMdef}.

As we discussed in section~\ref{t_y_system}, the leading eigenvalue of $T(u)$ in the r. h. s. of \eqref{BQTMlambdadef} generates the $T$-system \eqref{t_system}, cf. \eqref{eq:first:t}. In particular, both in the diagonal and non-diagonal case it is directly related to the first $y$-function $y_1(u)$ from \eqref{explicit_t}. Using then \eqref{BQTMlambdadef}, it is straightforward to write $y_1$ in terms of the leading eigenvalue $\Lambda_0$ of $\mathcal{T}$. Introducing the rotated function $\tilde{y}_1(\lambda)$ in \eqref{eq:rotated_y}, we have explicitly
\bea
1+\tilde{y}_1(\lambda)&=&\frac{\mathcal{N}(\lambda+i\eta/2)\mathcal{N}(\lambda-i\eta/2)}{\chi(\lambda)}\Lambda_0(\lambda+i\eta/2)\Lambda_0(\lambda-i\eta/2)
\nonumber \\ 
&\times& 
\left[\frac{\sin(\lambda+i(\beta/2N-\eta/2))\sin(\lambda-i(\beta/2N-\eta/2))}{\sin(\lambda+i(\beta/2N+\eta/2))\sin(\lambda-i(\beta/2N+\eta/2))}\right]^{2N}\,.
\label{y_lambda_relation}
\eea
Here, the function $\chi(\lambda)$ is simply obtained after rotation in the complex plane of the denominator of \eqref{explicit_t}. In the case of the N\'eel state, Eq.~\eqref{fdiagonal} yields $\chi(\lambda)=\chi_{\rm N}(\lambda)$, where
\be
\chi_{\rm N}(\lambda)=\frac{\sin(2\lambda+2i\eta)\sin(2\lambda-2i\eta)}{\sin(2\lambda+i\eta)\sin(2\lambda-i\eta)}\sin^{2}(\lambda+i\eta/2)\sin^{2}(\lambda-i\eta/2)\,,
\label{chidiag}
\ee
while in the case of tilted N\'eel state Eq.~\eqref{fnondiag} gives  $\chi(\lambda)=\chi_{\rm TN}(\lambda)$, where
\bea
\chi_{\rm TN}(\lambda)&=&
16 \kappa^4  \frac{\sin(2\lambda+2 i\eta)\sin(2\lambda-2 i\eta)}{\sin(2 \lambda+ i\eta)\sin(2\lambda-i \eta)}\nonumber\\
&\times & \left( \sin(\lambda+ i \eta/2)\sin(\lambda- i\eta/2)\cos(\lambda-i\zeta)\cos(\lambda+ i \zeta) \right)^2 
\,.
\label{chinondiag}
\eea
Analogously, the function $\mathcal{N}(\lambda)$ in \eqref{eq:norm} can be easily written in the case of the N\'eel state and tilted N\'eel states, for which we use the symbols $\mathcal{N}_{\rm N}(\lambda)$ and $\mathcal{N}_{\rm TN}(\lambda)$ respectively. Explicitly, they read
\bea
\mathcal{N}_{\rm N}(\lambda)=-\sin(\lambda+i\eta)\sin(\lambda-i\eta)-\sin^2(\lambda)\,,
\label{normdiag}
\eea 
and
\bea
\mathcal{N}_{\rm TN}(\lambda)&=& \kappa^2 \left[ -2 \cosh ^2(\zeta ) \cosh (2 \eta )\right.\nonumber\\
&+&\left.\cosh (2 \zeta ) (2 \cos (2 \lambda)-1)+4 \cosh (\eta ) \sin ^2(\lambda)+\cos (4 \lambda) \right] \,.
\label{normnondiag}
\eea

The functional relation \eqref{y_lambda_relation} can now be cast into the form of an integral equation, following the same prescription explained in section~\ref{t_y_system}. Importantly, we note that the function $\Lambda_0(\lambda)$ has no poles and no zeroes in the physical strip \eqref{physicalstrip}. We have verified numerically that this is the case, both for the N\'eel and tilted N\'eel state, for Trotter numbers $2N=2,4,6,8$. Then, the calculation is straightforward and here we only report the final result. We define the function $Y_1(\lambda)$ by
\be
1+ Y_1(\lambda ) =  \frac{\mathcal{N}(\lambda+i\eta/2)\mathcal{N}(\lambda-i\eta/2)}{\chi(\lambda)}  \,.
\label{definition_y1}
\ee
The explicit expressions for $Y_1(\lambda)$ for the N\'eel and tilted N\'eel states are immediately obtained by using \eqref{chidiag}, \eqref{normdiag} and \eqref{chinondiag}, \eqref{normnondiag} respectively. One then simply obtains
\be
\fl
\log\Lambda_0(\lambda)=
\left[s\ast \log\left(\frac{1+\tilde{y}_1}
{
1+Y_1
}
\right)\right](\lambda)+2N\sum_{k\in\mathbb{Z}}\frac{e^{-|k|\eta} e^{-2i k\lambda}}{\cosh(\eta k)}\frac{\sinh(\beta k/N)}{k}\,,
\,
\ee
where we employed the usual notation \eqref{convolution} for the convolution of two functions and where $s(\lambda)$ is given in \eqref{s_function}. We can now straightforwardly perform the Trotter limit. In particular, it is easy to see that for $N\to\infty$ one obtains
\be
\log \Lambda_0(\lambda)=\int_{-\pi/2}^{+\pi/2}\,d\mu\ s(\lambda-\mu)\left\{4\pi\beta a(\mu)+ \log\left[\frac{1+\tilde{y}_1(\mu)}{
1+Y_1(\lambda)
}\right]\right\}\,,
\ee
where we introduced
\be
a(\lambda)=\frac{1}{\pi}\frac{\sinh(\eta)}{\cosh(\eta)-\cos(2\lambda)}\,.
\label{a_function}
\ee

Using now \eqref{gLambda}, and rewriting $\beta=\beta_{w}$ in terms of $w$ [cf. \eqref{beta_parameter}] we arrive at the final expression for the dynamical free energy
\be
g(w)=\frac{1}{2}\int_{-\pi/2}^{+\pi/2}\,d\mu\ s(\mu)\left\{2\pi w J\sinh(\eta) a(\mu)+ \log\left[\frac{1+\tilde{y}_1(\mu)}{
1+Y_1(\mu)
}\right]\right\}\,.
\label{final_g}
\ee
This equation yields the value of the dynamical free energy once the first $y$-function $\tilde{y}_1(\lambda)$ is known. In the next section we discuss the solution of the TBA-like equations \eqref{TBA_diagonal} and \eqref{TBA_nondiagonal} and the subsequent numerical evaluation of \eqref{final_g} for real values of $w$. Finally, we consider the continuation to real times (namely imaginary values of $w$) and present our numerical results for the Loschmdit echo.

\section{The dynamical free energy and the Loschmidt echo}
\label{sec:resolution}

\subsection{The $\beta\to0$ limit: analytical solution}

In this section we show that equations \eqref{TBA_diagonal} and \eqref{TBA_nondiagonal} admit an analytical solution for $\beta=0$. More precisely, the solution for $\tilde{y}_1(\lambda)$ can be constructed analytically and the higher functions $\tilde{y}_n(\lambda)$ can be obtained recursively from the $Y-$system \eqref{y_system_lambda}.

From the definition \eqref{BQTMlambdadef}, using \eqref{eq:final_form} and after comparison with \eqref{eq:mathcal_t} we have
\be
\lim_{\beta\to 0}\mathcal{T}(\lambda)=\frac{\tensor[_{\mathcal{N}}]{\langle v^{+}(i\lambda) |T^{\rm QTM}(i\lambda)\otimes T^{\rm QTM}(-i\lambda) |v^{-}(i\lambda)\rangle}{_{\mathcal{N}}}}{\left[\sin\left(\lambda+i\eta\right)\sin\left(\lambda-i\eta\right)\right]^{2N}}\,,
\label{eq:mathcal_t_u}
\ee
where we introduced the normalized vectors
\bea
|v^{-}(i\lambda)\rangle_{\mathcal{N}}=\frac{|v^{-}(i\lambda)\rangle}{\sqrt{\mathcal{N}(\lambda)}}\,,\qquad \left(|v^{+}(i\lambda)\rangle_{\mathcal{N}}\right)^{\ast}=\frac{\left(|v^{+}(i\lambda)\rangle\right)^{\ast}}{\sqrt{\mathcal{N}(\lambda)}}\,,
\eea
and where $|v^{\pm}(u)\rangle$ are defined in \eqref{v1}, \eqref{v2}.
Analogously to \eqref{aux_eq} it is now easy to show the equivalence
\bea
\fl {\rm tr}\left\{\left[\tensor[_{\mathcal{N}}]{\langle v^{+}(i\lambda) |T^{\rm QTM}(i\lambda)\otimes T^{\rm QTM}(-i\lambda) |v^{-}(i\lambda)\rangle}{_{\mathcal{N}}}\right]^{L/2}\right\}\nonumber\\
=\langle \Psi^+_0 (\lambda) | \left[ t_{i\lambda,-i\lambda}(0) t_{i\lambda,-i\lambda}(-\eta) \right]^N  |  \Psi^-_0(\lambda) \rangle \,,
\label{aux2_eq}
\eea
where we defined
\bea
|\Psi^{-}_0(\lambda)\rangle=|v^{-}(i\lambda)\rangle_{\mathcal{N}}^{\otimes L/2}\,,\qquad |(\Psi^{+}_0(\lambda)\rangle)^{\ast}=\left[\left(|v^{+}(i\lambda)\rangle_{\mathcal{N}}\right)^{\ast}\right]^{\otimes L/2}\,.
\eea
Here we introduced the transfer matrix $t_{i\lambda,-i\lambda}$ acting in the original physical time direction, with inhomogeneitites in the physical space
\be
\xi_1=i\lambda,\ \xi_2=-i\lambda,\ \xi_3=i\lambda,\ \xi_4=-i\lambda,, \ldots\, ,
\ee
namely,
\be
\fl t_{i\lambda,-i\lambda}(w) = {\rm tr}_0 \left\{\mathcal{L}_L(w+i\lambda)\mathcal{L}_{L-1}(w-i\lambda) \ldots \mathcal{L}_2(w+i\lambda)\mathcal{L}_1(w-i\lambda)\right\} \,,
 \label{eq:transfermatrixt_dis}
\ee   
where $\mathcal{L}(w)$ is defined in \eqref{eq:mathcal_L}. Then, from \eqref{aux2_eq} it follows (in the limit $\beta \to 0$)
\be
\frac{\langle\Psi_0^{+}(\lambda)|\left[t_{i\lambda,-i\lambda}(0)t_{i\lambda,-i\lambda}(-\eta)\right]^N|\Psi^{-}_0(\lambda)\rangle}{\left[\sin(\lambda+i\eta)\sin(\lambda-i\eta)\right]^{NL}}=\tr\left\{\mathcal{T}(\lambda)\right\}^{L/2}\,.
\label{aux3_eq}
\ee
Since $\Lambda_0(0)$ is the leading eigenvalue of $\mathcal{T}(0)$ with a finite gap, for a neighborhood of $\lambda=0$, $\Lambda_0(\lambda)$ will continue to be the leading eigenvalue of $\mathcal{T}(\lambda)$. Hence, for small $|\lambda|$ one has
\be
\tr\left\{\mathcal{T}(\lambda)\right\}^{L/2}\simeq \Lambda_0(\lambda)^{L/2}\,.
\ee
On the other hand, for small $|\lambda|$ the following inversion relation holds
\be
\lim_{L\to\infty}\frac{t_{i\lambda,-i\lambda}(0)t_{i\lambda,-i\lambda}(-\eta)}{\left[\sin(\lambda+i\eta)\sin(\lambda-i\eta)\right]^{L}}=\mathrm{id}\,,
\ee
where ${\rm id}$ is the identity. More precisely, denoting with ${\rm id}_L$ the identity operator on a chain of $L$ sites, one can show that the Hilbert-Schmidt norm of the operator
\be 
\frac{t_{i\lambda,-i\lambda}(0)t_{i\lambda,-i\lambda}(-\eta)}{\left[\sin(\lambda+i\eta)\sin(\lambda-i\eta)\right]^{L}}-\mathrm{id}_{L}
\ee
is exponentially vanishing with the system size $L$, using techniques similar to those employed, for example, in \cite{prosen-14,ppsa-14,pv-16}. We refer the reader to these works for more details.

Putting everything together, \eqref{aux3_eq} yields
\be
\Lambda_0(\lambda)\equiv 1\,,
\label{lambda0}
\ee
for any finite Trotter number $N$. Note that \eqref{lambda0} has been derived under the assumption of small $|\lambda|$. However, assuming $\Lambda_0(\lambda)$ to be a meromorphic function in the complex plane, \eqref{lambda0} immediately implies $\Lambda_0(\lambda)\equiv 1$ for arbitrary values of $\lambda$ and at any finite Trotter number $N$.

Note that \eqref{lambda0} holds both in the diagonal and non-diagonal case and in principle it could be established by solving directly the Bethe equations in the limit $\beta\to 0$. In the case of the N\'eel state this calculation is easily done at finite Trotter number $N$, since all the Bethe roots approach $0$ as $\beta\to 0$. Conversely, this calculation is non-trivial in the non-diagonal case, as the extra Bethe roots approach non-zero values when $\beta\to 0$. In the case $2N=2$ we analytically solved the Bethe equations and showed that \eqref{lambda0} is verified for $\beta\to 0$ as we report in \ref{app:bethe_eq}, while for higher values of $N$ we verified this numerically.

Using now \eqref{lambda0} Eq.~\eqref{y_lambda_relation} simply yields
\be
\lim_{\beta\to 0}\tilde{y}_1(\lambda)=Y_1(\lambda)\,,
\label{solution_y1}
\ee
where $Y_1(\lambda)$ has been defined in \eqref{definition_y1}. It is now immediate to generate the analytic solution for $\tilde{y}_n(\lambda)$ as
\be
\lim_{\beta\to 0}\tilde{y}_n(\lambda)=Y_n(\lambda)
\label{solution_yn}
\ee
where $Y_n(\lambda)$ is defined recursively from $Y_1(\lambda)$ using the $Y$-system
\be
Y_j\left(\lambda+i\frac{\eta}{2}\right)Y_j\left(\lambda-i\frac{\eta}{2}\right)=\left[1+Y_{j+1}\left(\lambda\right)\right]\left[1+Y_{j-1}\left(\lambda\right)\right]\,,
\label{y_system_capitol_y}
\ee
where $Y_0(\lambda)=0$.

Equations \eqref{solution_y1}, \eqref{solution_yn} provide the analytic solution of the equations \eqref{TBA_diagonal} and \eqref{TBA_nondiagonal} for $\beta=0$. As we already stressed in the last section, the explicit expressions for the N\'eel and tilted N\'eel states are immediately obtained by means of \eqref{chidiag}, \eqref{normdiag} and \eqref{chinondiag}, \eqref{normnondiag} respectively. We next address the numerical solution for general value of $\beta$.

\subsection{General $\beta$: numerical evaluation}

We now discuss the numerical solution of the TBA equations \eqref{TBA_diagonal} and \eqref{TBA_nondiagonal} for general $\beta$. First, the infinite sets of equations have to be truncated to a finite number $n_{\rm max}$. One then immediately notes that these equations alone do not completely constrain the functions $\tilde{y}_n(\lambda)$, as one should also provide an asymptotic condition on the behavior of $\tilde{y}_n(\lambda)$ for $n\to\infty$.

For $\beta=0$ the analytic solution of \eqref{TBA_diagonal} and \eqref{TBA_nondiagonal} was provided in the previous section. In particular, the analytical knowledge of $\tilde{y}_1(\lambda)$ enables us to generate through \eqref{y_system_capitol_y} the whole set of functions $\tilde{y}_n(\lambda)$. Then, one can simply study the behavior of $\tilde{y}_n(\lambda)$ for large $n$ at $\beta=0$. As expected, we observe that the asymptotic behavior strongly depend on the initial state considered, and a separate analysis for each state has to be performed.

As we already observed, and will discuss systematically in section~\ref{sec:qam}, the TBA equations \eqref{TBA_diagonal} corresponding to the N\'eel state  for $\beta=0$ coincide with the generalized TBA equations derived in \cite{wdbf-14,wdbf-14_2,pmwk-14} by means of the quench action method. In particular, they share the same solution and the large-$n$ behavior is found to be \cite{wdbf-14,wdbf-14_2}
\bea
\tilde{y}_{2n}(\lambda)\Big|_{\beta=0}&\sim & h_1(\lambda)\,,\nonumber\\
\tilde{y}_{2n+1}(\lambda)\Big|_{\beta=0}&\sim & h_2(\lambda)\,,
\label{as_condition_1}
\eea
where $h_1(\lambda)$, $h_2(\lambda)$, are non-trivial functions.

Instead, in the case of the tilted N\'eel state \eqref{tilted_neel} with tilting angle $\vartheta\neq 0$ we find
\bea
\tilde{y}_{2n}(\lambda)\Big|_{\beta=0}&\sim & (2n)^4g_1^{\vartheta}(\lambda)\,,\nonumber\\
\tilde{y}_{2n+1}(\lambda)\Big|_{\beta=0}&\sim &(2n+1)^4 g_2^{\vartheta}(\lambda)\,,
\label{as_condition_2}
\eea
where again $g_1(\lambda)^{\vartheta}$ and $g_2(\lambda)^{\vartheta}$ are non-trivial $\vartheta $-dependent functions.

The asymptotic conditions \eqref{as_condition_1} and \eqref{as_condition_2} can be easily be implemented in the numerical solution of the truncated system of $n_{\rm max}$ equations by setting respectively
\be
\tilde{y}_{n_{\rm max}+1}(\lambda)=\tilde{y}_{n_{\rm max}-1}(\lambda)   \qquad \mbox{(N\'eel state)} \,,
\label{closure_condition_diagonal}
\ee
and
\be
\tilde{y}_{n_{\rm max}+1}(\lambda)=\left(1+\frac{2}{n-1}\right)^4\tilde{y}_{n_{\rm max}-1}(\lambda) 
 \qquad \mbox{(tilted N\'eel state)}
\,.
\label{closure_condition_nondiagonal}
\ee
In fact, the resulting system of $n_{\rm max}$ equations can then be solved by an iterative procedure. As a consistency check, we have verified that numerical solution of \eqref{TBA_diagonal} and \eqref{TBA_nondiagonal} for $\beta=0$ under the truncation conditions \eqref{closure_condition_diagonal} and \eqref{closure_condition_nondiagonal} yields correctly the analytical solutions derived in the previous section.

\begin{figure}
\hspace{-0.6cm}
\includegraphics[scale=0.75]{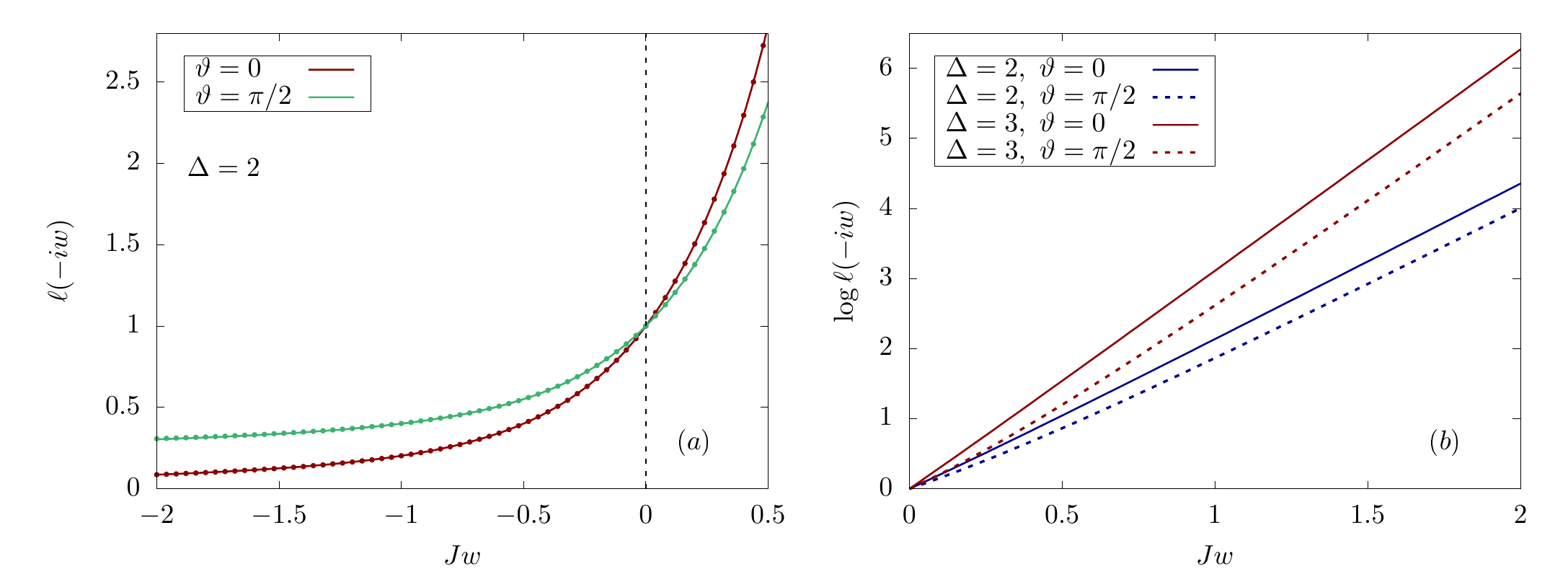}
\caption{Loschmidt echo per site at imaginary times for the initial tilted N\'eel state \eqref{tilted_neel}. ($a$)~:~The plot shows the analytical result from numerical evaluation of \eqref{final_g} (solid lines) for $\Delta=2$ and different tilting angles, together with exact diagonalization data for a system of $L=16$ sites (dots). The dashed vertical line is a guide for the eye corresponding to $w$=0. ($b$)~:~Logarithmic plot for different values of $\Delta$ and tilting angle $\vartheta$, showing exponential behavior of the Loschmidt echo per site at large imaginary times, as computed from \eqref{final_g}.}
\label{fig:real_beta}
\end{figure}

For $\beta\neq 0$ the asymptotic condition for large $n$ of $\tilde{y}_n(\lambda)$ is not known and a priori might depend non-trivially on $\beta$. In the thermal case, no dependence on the temperature arises in the case of zero magnetization, where no magnetic field is present \cite{takahashi-99}. In analogy with this case, we will assume the validity of \eqref{closure_condition_diagonal}, and \eqref{closure_condition_nondiagonal} also for $\beta\neq 0$. It should be stressed that this is a priori a non-trivial assumption, the validity of which is justified a posteriori, given the excellent numerical agreement with independent numerical methods. As we will see, in the case of the N\'eel state a direct check of \eqref{closure_condition_diagonal} for $\beta\neq 0$ is also possible using the coupled version of the integral equations \eqref{TBA_diagonal}, cf. section~\ref{sec:qam}. Finally, note that a $\beta$-dependence might arise for generic states with non-zero magnetization such as tilted ferromagnets, for which a more systematic analysis might be needed.

Truncation of \eqref{TBA_diagonal}, \eqref{TBA_nondiagonal} using
\eqref{closure_condition_diagonal} and
\eqref{closure_condition_nondiagonal} results in a finite set of
integral equations whose numerical solution appears to be stable. In
particular, one observes that increasing $n_{\rm max}$ the function
$\tilde{y}_1(\lambda)$ rapidly converges to a well-defined function
for $\beta\neq 0$ (in practice excellent numerical accuracy is obtained using $n_{\rm max} \sim 20$). Accordingly, \eqref{final_g} immediately yields our
 prediction for the dynamical free energy $g(w)$.  In our practical
 examples we chose the tilting angle $\vartheta$ such that the
 parameter $\zeta$ is a multiple of $\eta/2$ (this includes the case
 of $\zeta=0$, $\vartheta=\pi/2$). This way all source
 terms in the TBA equations can be written using Jacobi functions,
 cf. \eqref{dn-functions} and \eqref{jacobi}.
  
The final result is shown in Fig.~\ref{fig:real_beta}, where we plot the Loschmidt echo per site at imaginary times. First, we can see that our solution is in excellent agreement with exact diagonalization data for a finite chain of $L=16$ sites. 
As expected, we see that the dynamical free energy displays a monotonic behavior. Furthermore for large positive $w$ an exponential increase is observed, as we can clearly see from the figure. In the next section we will discuss the more interesting picture emerging for imaginary $\beta$, namely real times.

\subsection{Continuation to real times: the Loschmidt echo}

From the derivation of the last sections, the TBA-like equations \eqref{TBA_diagonal}, \eqref{TBA_nondiagonal} are expected to continue to provide the correct results also in the case where $\beta$ is a complex parameter, provided that two conditions remain valid:
\begin{enumerate}
\item First, the analytic structure of the $Y$-functions $\tilde{y}_n(\lambda)$ should remain the same as in the case of $\beta$ real. This means that no additional poles or zeroes of $\tilde{y}_n(\lambda)$ should enter into the physical strip, beyond those already described in section~\ref{sec:TBA_equations}.\label{condition_1}
\item The second condition is that the leading eigenvalue of the boundary transfer matrix \eqref{eq:mathcal_t} should not display a crossing with subleading eigenvalues. \label{condition_2}
\end{enumerate}

\begin{figure}
\hspace{-0.6cm}
\includegraphics[scale=0.75]{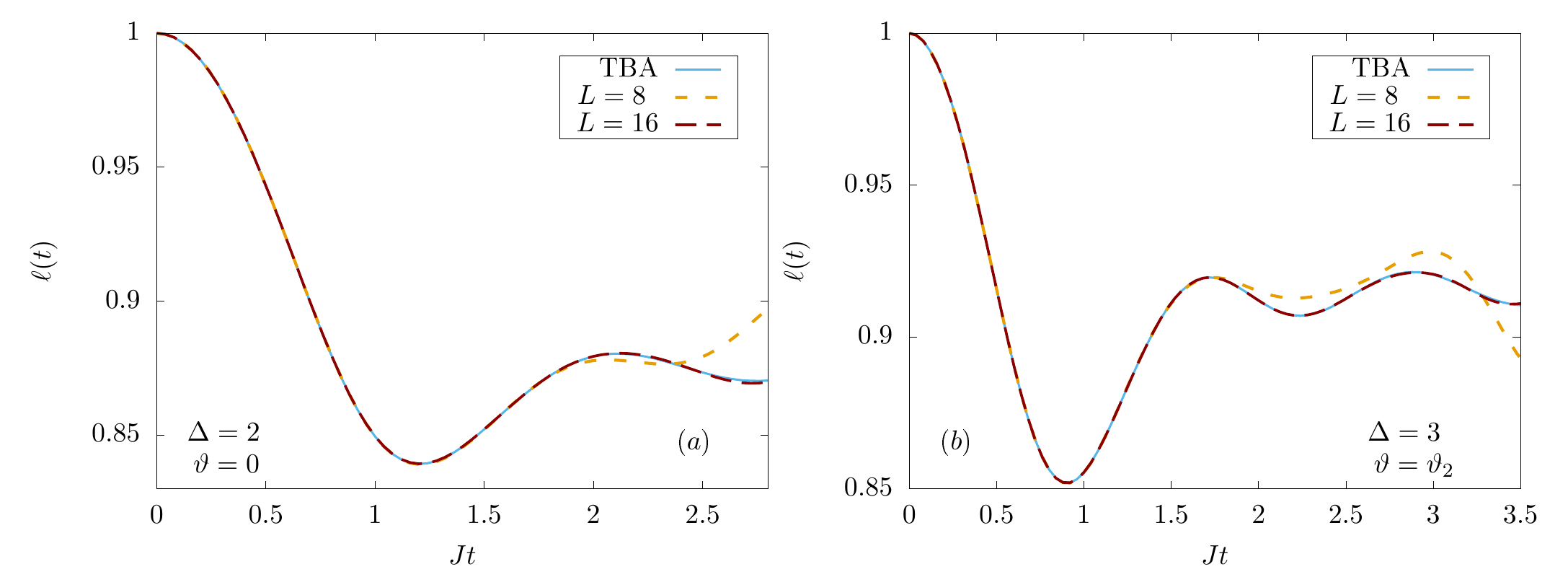}
\caption{Loschmidt echo per site \eqref{eq:losch} for the tilted N\'eel state \eqref{tilted_neel}. The plots correspond to ($a$)~$\Delta=2$ and tilting angle $\vartheta=0$ and ($b$)~$\Delta=3$ and tilting angle $\vartheta_2=2{\rm arctan}(e^{-\eta})\simeq 0.11\pi $. Exact diagonalization data for increasing system sizes (dashed lines) are shown together with the analytical predictions by \eqref{final_g} (solid lines). }
\label{fig:check_vs_ED}
\end{figure}

Note that these conditions are not always necessarily true for arbitrary complex values of $\beta$. Since these conditions are verified for $\beta$ real, they are
expected to hold also for complex $\beta$
in some neighbourhood of the real axis.
In fact, we considered the solution of
\eqref{TBA_diagonal}, \eqref{TBA_nondiagonal} for purely imaginary
values of $\beta$ (corresponding to real times) and found that for
small times the numerical solution was always stable for all the
values of the parameters that we chose. We compared the corresponding
predictions obtained from \eqref{final_g} to exact diagonalization
data and found excellent agreement as shown in
Fig.~\ref{fig:check_vs_ED}. This provides a non-trivial check on the
validity of our results. 

\begin{figure}
\centering
\includegraphics[scale=1.05]{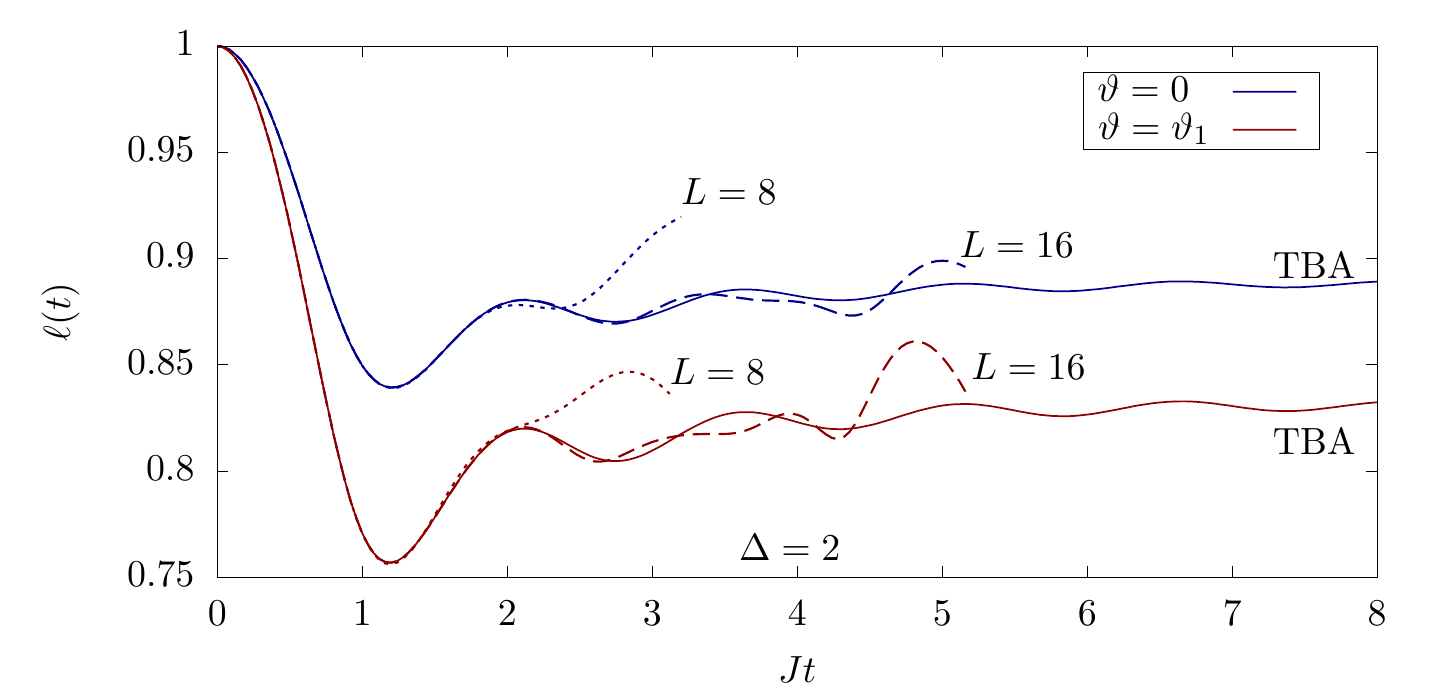}
\caption{Loschmidt echo per site \eqref{eq:losch} for the tilted N\'eel state \eqref{tilted_neel} for $\Delta=2$ and tilting angles $\vartheta=0$, $\vartheta_1=2{\rm arctan}(e^{-\eta})\simeq 0.17\pi $. The analytical predictions by \eqref{final_g} (solid lines) are shown, together with exact diagonalization data for systems of $L=8$, $L=16$ sites.}
\label{fig:real_times_D2}
\end{figure}

\begin{figure}
\centering
\includegraphics[scale=1.05]{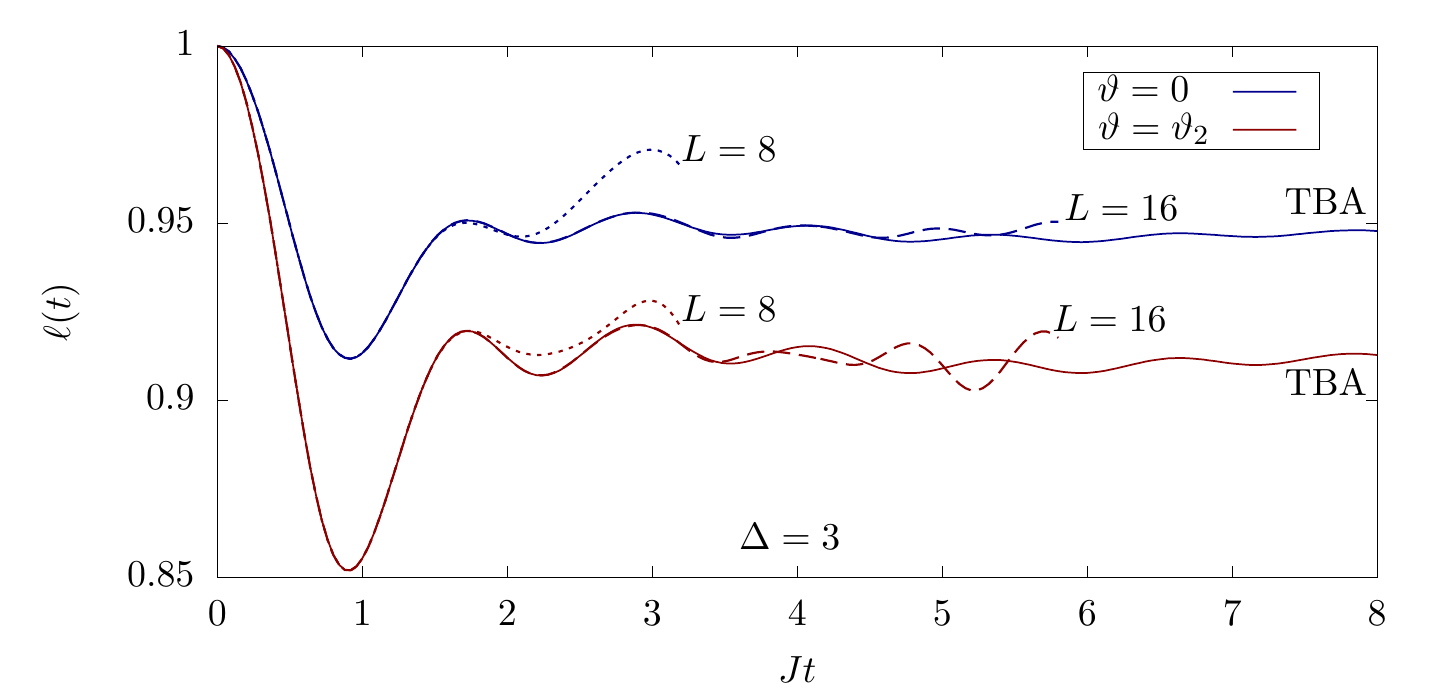}
\caption{Loschmidt echo per site \eqref{eq:losch} for the tilted N\'eel state \eqref{tilted_neel} for $\Delta=3$ and tilting angles $\vartheta=0$, $\vartheta_1=2{\rm arctan}(e^{-\eta})\simeq 0.11\pi $. The analytical predictions by \eqref{final_g} (solid lines) are shown, together with exact diagonalization data for a system of $L=8$, $L=16$ sites.}
\label{fig:real_times_D3}
\end{figure}

When the times are increased, one should test the validity of the two
conditions above. If they are valid, then
\eqref{TBA_diagonal} and \eqref{TBA_nondiagonal} can be used to
calculate the Loschmidt echo per site through \eqref{final_g}. The
check of the validity of the conditions \eqref{condition_1},
\eqref{condition_2} will be discussed shortly, while we present in
Figs.~\ref{fig:real_times_D2}, \ref{fig:real_times_D3} two examples
where these are verified and \eqref{TBA_diagonal} and
\eqref{TBA_nondiagonal} can be used to compute $\ell(t)$ up to very
large times. Our analytical predictions are shown together with exact
diagonalization data for systems of $L=8,16$ sites. These are seen to
be in excellent agreement for small times, while as expected
deviations occur at larger times due to the fact that finite size
effects become dominant. We see that in contrast to finite size
results, the amplitude of the oscillation of $\ell(t)$ in the
thermodynamic limit appears to be rapidly damped with time.  

We now turn to testing the conditions \eqref{condition_1},
\eqref{condition_2} for imaginary $\beta$ (real times). The validity of condition
\eqref{condition_1} can be checked straightforwardly. Indeed, as an
additional pole or zero of $\tilde{y}_n(\lambda)$ comes close and
enters into the physical strip at time $t_0$, one of the functions
$\log(1+\tilde{y}_{n+1}(\lambda))$, $\log(1+\tilde{y}_{n-1}(\lambda))$ exhibits the
emergence of a singularity for real $\lambda$, due to the $Y$-system
relation \eqref{y_system_lambda}. If we don't modify the TBA equations
and solve the same equations \eqref{TBA_diagonal} and
\eqref{TBA_nondiagonal} also for $t> t_0$, the
corresponding numerical curve for the Loschmidt echo displays a
fictitious non-analyticity point. 

\begin{figure}
\hspace{-0.6cm}
\includegraphics[scale=0.75]{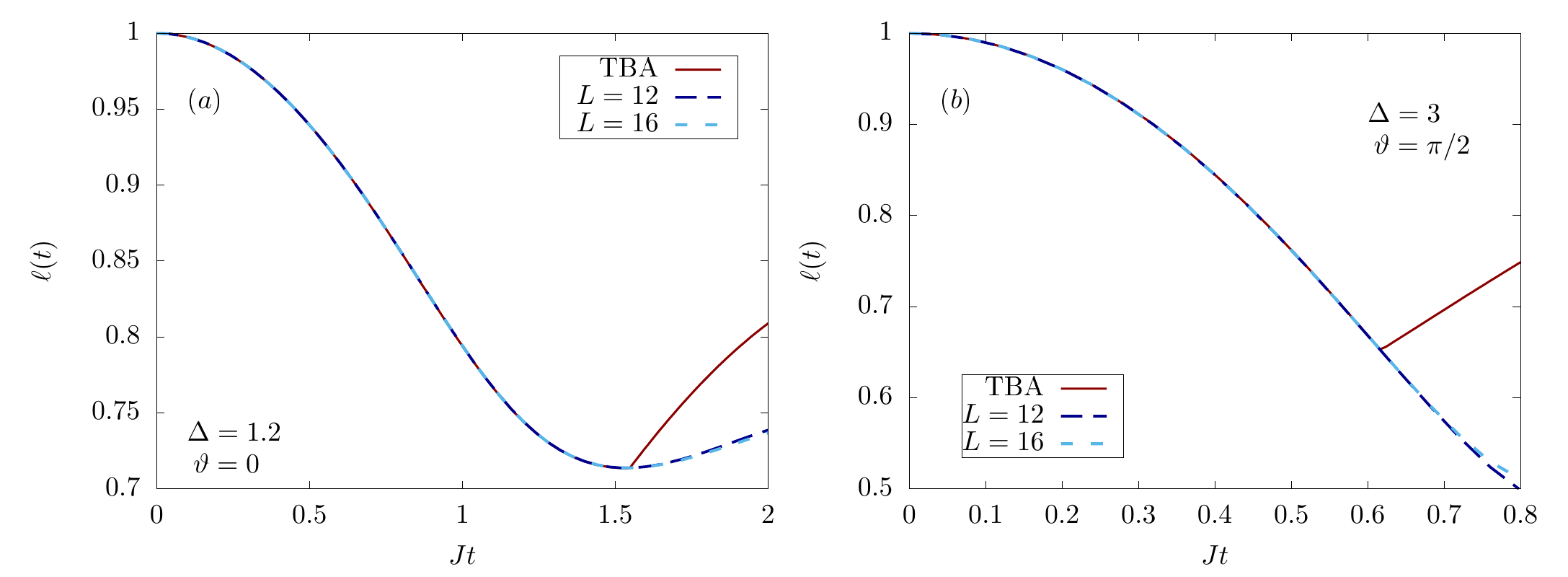}
\caption{Loschmidt echo per site \eqref{eq:losch} for the tilted N\'eel state \eqref{tilted_neel}. The plots correspond to ($a$) $\Delta=1.2$ and tilting angle $\vartheta=0$ and ($b$) $\Delta=3$ and tilting angle $\vartheta=\pi/2$. Exact diagonalization data for increasing system sizes (dashed lines) are shown together with the analytical predictions by \eqref{final_g} (solid lines). The time $t_0$ at which the integral equations \eqref{TBA_diagonal} and \eqref{TBA_nondiagonal} stop to be valid is clearly visible as a fictitious point of non-analyticity.}
\label{fig:analytic_break}
\end{figure}

We verified that these additional singularities do in fact arise, depending
on the parameters of the initial states. In particular, this is the
case for $\Delta\sim 1$ and large tilting angles, as clearly shown in
Fig.~\ref{fig:analytic_break}. We see that the numerical solution of
\eqref{TBA_diagonal} and \eqref{TBA_nondiagonal} displays a break with
respect to exact diagonalization data. This fictitious point of
non-analyticity corresponds to the entering of a zero of
$\tilde{y}_2(\lambda)$ inside the physical strip. We have verified
this by looking at the analytic structure of $\tilde{y}_2(\lambda)$
(as obtained from the QTM method) at
finite Trotter number $2N=4$, $2N=6$. In these cases an additional
zero of $\tilde{y}_2(\lambda)$ is clearly seen entering the physical
strip at a time which depends very weakly on $N$ and is consistent
with the fictitious point of non-analyticity of
Fig.~\ref{fig:analytic_break}. Alternatively, it can be seen from the
TBA equations that a zero of $(1+\tilde y_1(\lambda))$ crosses the real line.

When such additional poles and zeroes enter the physical strip, the TBA equations \eqref{TBA_diagonal} and
\eqref{TBA_nondiagonal} have to be modified by inserting additional
source terms corresponding to the new singularities. This can in
principle be done with the same techniques as those explained in
section~\ref{sec:TBA_equations}.
In particular, the results are expected to take the form of 
"excited state TBA equations", similar to those encountered in the framework of integrable quantum field
theory \cite{dt-96}.
However, a detailed analysis of
this problem
goes beyond the
scope of the present work.

The test of condition \eqref{condition_2} and the determination of
the precise time of its failure are more subtle. In the cases
displayed in Figs.~\ref{fig:real_times_D2}, \ref{fig:real_times_D3},
namely $\Delta=2,3$ and tilting angle $\vartheta$ not too large, we
studied the spectrum of the transfer matrix \eqref{eq:mathcal_t} for
finite Trotter numbers $2N =4, 6, 8$. This analysis indicated that in
these cases either no crossing arise at all or one has crossing at
very large times (much larger than those displayed in
Fig.~\ref{fig:real_times_D2}, \ref{fig:real_times_D3}). Conversely,
when the anisotropy $\Delta$ is close to one or for large tilting
angles as in Fig.~\ref{fig:analytic_break} we can clearly observe a
crossing between the leading and subleading eigenvalue at finite
Trotter number $2N=4,6$. The crossing happens at a time $t_1$ which
weakly depends on $N$ and which is larger than the fictitious
non-analyticity point shown in Fig. \ref{fig:analytic_break}.
 
In this case care must be taken, as the treatment of the previous
section is valid only up to the crossing time $t_1$. In order to
determine the precise location of the crossing in the Trotter
limit and the time evolution afterwards a more sophisticated treatment
is required. In particular, one should compute the first excited
states of the boundary transfer matrix \eqref{eq:mathcal_t} in the
Trotter limit, by generalizing the method employed here for the
leading eigenvalue. The excited eigenvalues will display a different
analytical structure and hence will correspond to ``excited state TBA
equations'', which were encountered earlier in integrable quantum field theory
\cite{dt-96} or in spin chains describing the exponential decay of
correlations \cite{klum-93}.
By solving these, one has access to higher
excited states in the Trotter limit, from which one can compute
the time of the crossing and subsequent time evolution, by
continuously following at each step the leading eigenvalue. Note that
for each crossing a genuine (i.e. not fictitious) point of
non-analyticity will be encountered. As we already mentioned in the
introduction, such non-analyticities can arise also for quenches
within the same quantum phase \cite{fagotti2-13,as-14,vd-14}.  The
computation of the full Loschmidt echo in the presence of crossings
goes once again beyond the scope of this work and remains an
interesting direction to be addressed in the future. 

\section{From the quantum transfer matrix to the quench action}\label{sec:qam}

In this section we finally draw a parallel between the quantum
transfer matrix approach and the quench action method, which is
constructed in the language of the thermodynamic Bethe ansatz (TBA). An
analogous picture emerges also in the thermal case, where the
equivalence between the TBA and the quantum
transfer matrix formalism is an established result \cite{tsk-01}. In
the following, we first briefly review some aspects of the quench
action method, referring to the literature for a more complete
treatment \cite{caux-16}. We stress that the calculations of this
section are limited to imaginary times, when the quench action is
strictly real. At the end of this section we give a few comments on
the case of real times.

The central idea of the thermodynamic Bethe ansatz approach (both in
the thermal case and in the quench situations) is to 
investigate the original physical spin chain in the thermodynamic
limit, and to find the Bethe states which populate the system.
Within this formalism, eigenstates are
described by rapidity distribution functions
$\{\rho_n(\lambda)\}_{n=1}^{\infty}$. The function $\rho_n(\lambda)$
corresponds to bound states of $n$ quasi-particles (magnons) and
generalize the concept of momentum distribution functions for free
systems. In the same way, to each eigenstate is also associated a set
of hole rapidity distribution functions $\rho^h_n(\lambda)$, which are
analogous to the well-known hole distribution functions for free Fermi
gases. In interacting models, rapidity and hole distribution functions
are non-trivially related by the thermodynamic version of the Bethe
equations 
\be
\rho_{m}(\lambda)+\rho^{h}_{m}(\lambda)=a_m(\lambda)-\sum_{n=1}^{\infty}\left[a_{mn}\ast\rho_n\right](\lambda)\,,
\label{eq:thermo_bethe}
\ee
where we employed the notation \eqref{convolution} for the convolution and where
\be
\fl a_{mn}(\lambda)=(1-\delta_{mn})a_{|m-n|}(\lambda)+2a_{|m-n|}(\lambda)+\ldots +2a_{m+n-2}(\lambda)+a_{m+n}(\lambda)\,,
\ee
\be
a_n(\lambda)=\frac{1}{\pi} \frac{\sinh( n\eta)}{\cosh (n \eta) - \cos( 2 \lambda)}\,.
\label{def:a_function}
\ee
In the following, it is also convenient to introduce the ratios
\be
\eta_n(\lambda)=\frac{\rho_n^{h}(\lambda)}{\rho_n(\lambda)}\,.
\label{eta_function}
\ee

The functions $\rho_n(\lambda)$ in principle give access to the computation of all thermodynamical properties of the corresponding state. In particular, the energy per unit length is given as
\be
e\left[\{\rho_{n}\}_{n=1}^{\infty}\right]=\sum_{n=1}^{\infty}\int_{-\pi/2}^{\pi/2}{\rm d}\lambda\,\ \rho_n(\lambda)\varepsilon_n(\lambda)\,,
\label{energy}
\ee
where
\be
\varepsilon_n(\lambda)=-J\frac{\sinh(\eta)\sinh(n\eta)}{\cosh(n\eta)-\cos(2\lambda)}\,.
\ee
Note that the same set $\{\rho_n(\lambda)\}_{n=1}^\infty$ actually corresponds to many eigenstates, which can be interpreted as different microscopic realizations of the same macrostate. In the following we denote with $| \brho \rangle_L$ one of the possible eigenstates of the finite system which is described by the set $\{\rho_n(\lambda)\}_{n=1}^{\infty}$ in the thermodynamic limit.

In the computation of physical quantities following a quench, a building block is provided by the overlaps between the initial state $|\Psi_0\rangle$ and the eigenstates of the system (the Bethe states). The first essential step in the application of the quench action method relies then in the possibility of defining a functional $S_{\rm Q}$ such that for $L\to \infty$
\be
|\langle \Psi_0|\brho\rangle_{L}|\simeq \exp\left\{-LS_{\rm Q}\left[\{\rho_n(\lambda)\}_{n=1}^{\infty}\right]\right\}\,.
\label{eq:overlap}
\ee
It is important to stress that the validity of \eqref{eq:overlap} is a priori non-trivial, due to the fact that $S_{\rm Q}$ only depends on the set $\{\rho_n(\lambda)\}_{n=1}^{\infty}$ and not on the details of the state $|\brho\rangle_L$ chosen. In fact, the determination of the functional $S_{\rm Q}$ still represents the major technical bottle-neck to the application of the quench action approach for generic initial states $|\Psi_0\rangle$ \cite{dwbc-14, pozsgay-14,bdwc-14,pc-14}.

In this section we focus on the symmetric N\'eel state 
\be
|N\rangle_s=\frac{1}{\sqrt{2}}\left(|\uparrow\downarrow\ldots\uparrow\downarrow\rangle + |\downarrow\uparrow\ldots\downarrow\uparrow\rangle\right)\,.
\label{eq:symmetric_neel}
\ee
In this case, the overlaps with the Bethe states and the functional $S_{\rm Q}$ in \eqref{eq:overlap} have recently been computed \cite{wdbf-14,wdbf-14_2,pmwk-14, pozsgay-14,bdwc-14}. In particular, the latter has the simple expression \cite{wdbf-14,wdbf-14_2,pmwk-14}
\be
S_{\rm Q}\left[\{\rho_n(\lambda)\}_{n=1}^{\infty}\right]=-\frac{1}{4}\sum_{n=1}^{\infty}\int_{-\pi/2}^{\pi/2}{\rm d}\lambda \rho_{n}(\lambda)\log W_{n}(\lambda)\,,
\label{sq_expression}
\ee
where
\bea
 W_n (\lambda)& =& \frac{1}{2^{n+1} \sin^{2}2\lambda } \frac{ \cosh n \eta - \cos 2\lambda}{\cosh n \eta + \cos 2\lambda}\nonumber\\
&\times & \prod_{j=1}^{\frac{n-1}{2}} \left( \frac{ \cosh \left[(2j-1) \eta\right] - \cos 2\lambda}{(\cosh \left[(2j-1) \eta\right] + \cos 2\lambda) (\cosh 4\eta j - \cos 4\lambda )}   \right)^{2}\,,
\eea
if $n$ odd, and 
\bea
 W_n (\lambda)& = &\frac{ \tan^{2}\lambda}{2^{n} } \frac{ \cosh n \eta - \cos 2\lambda }{\cosh n \eta + \cos 2\lambda} \frac{1}{\prod_{j=1}^{\frac{n}{2}} \left(\cosh \left[2(2j-1)\eta\right] - \cos 4\lambda \right)^{2} } \nonumber\\
 &\times & \prod_{j=1}^{\frac{n-2}{2}} \left( \frac{\cosh 2j\eta - \cos 2\lambda }{\cosh 2j\eta + \cos 2\lambda }\right)^{2}\,,
\eea
if $n$ even.

Note that a priori the Loschdmit echo per site corresponding to the
symmetric state \eqref{eq:symmetric_neel} and the N\'eel state
\eqref{eq:neel} can be different, even in the thermodynamic
limit. This was for example explicitly shown by exact calculations in
\cite{as-14} for quenches to the free $XX$ spin chain. In that case,
the real-time Loschmidt echo per site in the two cases coincided up to
a maximum time $t_0$, after which they displayed different points of
non-analyticity and hence different behavior. On the other hand, the
same calculations showed that, for that quench, the dynamical free
energy \eqref{eq:g_en_function} at imaginary times is the same for the
two states. The physical reason for the difference at real times is
that for certain regimes in $t$ the transition amplitude 
\be
\langle\uparrow\downarrow\ldots\uparrow\downarrow|
e^{-iHt}
 |\downarrow\uparrow\ldots\downarrow\uparrow\rangle
\label{eq:transition}
\ee
can be bigger than the return rate of the pure N\'eel state \cite{as-14}.

In this section we will compute the dynamical free energy for the
state \eqref{eq:symmetric_neel} at imaginary times, by means of the
quench action approach. In analogy to the case of quenches to the free
$XX$ spin chain, and in accordance with numerical evidence for finite
system sizes, our calculation will show that the final result
coincides with the dynamical free energy of the N\'eel state
\eqref{eq:neel}. By comparing the solutions obtained by
the QTM and QA methods
we will draw an explicit correspondence
between the two approaches. 
 
Once the functional $S_{\rm Q}$ in \eqref{eq:overlap} is known, the application of the quench action approach is straightforward. By making use of the formal resolution of the identity in terms of rapidity distribution functions \cite{takahashi-99}, it is immediate to write
\be
\langle\Psi_0|e^{-wH}|\Psi_0\rangle=\int\mathcal{D}\brho \exp\left\{-\left(w e\left[\brho\right]+2S_{\rm Q}\left[\brho\right]-S_{\rm YY}\left[\brho\right]\right)L\right\}\,,
\label{eq:resolution_identity}
\ee
where we denoted with $\brho$ the set $\{\rho_n(\lambda)\}_{n=1}^{\infty}$ and where $e[\brho]$ is given in \eqref{energy}. Here $S_{\rm YY}$ is the so called Yang-Yang entropy and for parity symmetric states such as the N\'eel state, the latter has an additional factor $1/2$ with respect to the thermal case \cite{dwbc-14}. Thus, it reads
\bea
S_{\rm YY}\left[\left\{\rho_{n=1}^{\infty}(\lambda)\right\}\right](\lambda)&=&\frac{1}{2}\sum_{n=1}^{\infty}\int_{-\pi/2}^{\pi/2}{\rm d}\lambda \left\{\rho_n(\lambda)\log\left[1+\eta_n(\lambda)\right]\right.\nonumber\\
&+&\left.\rho^h_n(\lambda)\log\left[1+\eta^{-1}_n(\lambda)\right]\right\}\,.
\eea

The functional integral \eqref{eq:resolution_identity} can now be evaluated at its saddle-point. By following a standard derivation, similar to the one detailed in \cite{wdbf-14,wdbf-14_2,pmwk-14}, the saddle-point equations are immediately derived and read 
\bea
\log[\eta_{n}(\lambda)] & =& -2\pi J w\sinh(\eta) a_n(\lambda) - 2nh-\log W_{n}(\lambda)\nonumber\\
 &+& \sum_{m=1}^{\infty} \left[ a_{nm} \ast \log \left( 1 + \eta_{m}^{-1} \right)\right] (\lambda)\,,
\label{eq:saddle-point}
\eea
where the parameter $h$ is a Lagrange multiplier fixing the total magnetization. By plugging the solution of \eqref{eq:saddle-point} into the Bethe equations \eqref{eq:thermo_bethe} one finally obtains the saddle-point distribution functions $\brho^{\ast}=\{\rho^{\ast}_n\}_{n=1}^{\infty}$. The final result for the dynamical free energy then reads
\be
g(w)=-we[\brho^{\ast}]-2S_{\rm Q}[\brho^{\ast}]+S_{\rm YY}[\brho^{\ast}]\,.
\label{eq:temp_g}
\ee

Applying now the standard techniques of thermodynamic Bethe ansatz \cite{takahashi-99}, the equations \eqref{eq:saddle-point} can be cast in the partially decoupled form
\bea
\log[\eta_1(\lambda)]&=&-4\pi\beta_{w} s(\lambda)+d_1(\lambda)+\left[s\ast\log\left(1+\eta_2\right)\right](\lambda)\,,\\
\log[\eta_n(\lambda)]&=&d_n(\lambda)+\left[s\ast\left\{\log\left(1+\eta_{n-1}\right)+\log\left(1+\eta_{n+1}\right)\right\}\right](\lambda)\,,
\label{eta_TBA}
\eea
where we introduced the parameter $\beta_{w}$ as in \eqref{beta_parameter}, while the functions $s(\lambda)$ and $d_n(\lambda)$ are defined in \eqref{s_function} and \eqref{d_n_function} respectively. Finally, making use of \eqref{eta_TBA} it is shown in \ref{app:dyn_free_en} that the expression in \eqref{eq:temp_g} can be recast in the form
\be
g(w)=\frac{1}{2}\int_{-\pi/2}^{+\pi/2}\,d\mu\ s(\mu)\left\{2\pi w J\sinh(\eta) a_1(\mu)+ \log\left[\frac{1+\eta_1(\mu)}{
1+Y_1(\mu)
}\right]\right\}\,,
\label{QA_final_g}
\ee
where $Y_1(\mu)$ is given in \eqref{definition_y1}.

Remarkably, we can now immediately see that the result by the quantum transfer matrix is recovered. In particular, the function $\eta_n(\lambda)$ and $\tilde{y}_n(\lambda)$ are the solution of the same set of equations, leading to the identification
\be
\eta^{\beta}_n(\lambda)=\tilde{y}^{\beta}_n(\lambda)\,\qquad \beta\in\mathbb{R}\,,
\label{identification}
\ee
where we have made explicit the $\beta$-dependence. This equation unveils the direct link between the quantum transfer matrix and quench action methods. In particular, the Bethe ansatz rapidity distribution functions $\eta_n(\lambda)$ defined in \eqref{eta_function} are identified with the $y$-functions associated with the fusion of boundary transfer matrices. As we already mentioned, this extends the picture already established in the thermal case \cite{tsk-01}.

Going further, we can consider \eqref{identification} when $\beta_w\to
0$, namely $w\to 0$. Comparing to \eqref{eq:resolution_identity}, we
see that in the language of the quench action approach the
saddle-point distribution $\brho^\ast$ corresponds, when $w\to 0$, to
the representative eigenstate for the initial state $|\Psi_0\rangle$
\cite{caux-16}. This also follows from the decomposition of the
initial state as
\begin{equation}
  1=\langle\Psi_0|\Psi_0\rangle=\sum_{n} \Big|\langle\Psi_0|n\rangle\Big|^2\,.
\end{equation}
The l.h.s. gives us the seemingly trivial result that the Loschmidt
amplitude at $w=0$ is zero, whereas the r.h.s. leads to the standard saddle
point evaluation of the same quantity using the quench action method.

Hence, assuming the identification \eqref{identification} we can obtain the functions $\eta_n(\lambda)$ for the representative eigenstate directly from the quantum transfer matrix construction, without any reference to the quench action derivation. Indeed, in the previous sections we obtained the analytical solution for $\tilde{y}_1(\lambda)$ at $\beta=0$, cf. \eqref{solution_y1}. This holds for generic two-site product states, for which the overlaps with the Bethe states are not yet known. Considering in particular the example of the tilted N\'eel state, we arrive immediately at the result
\be
\eta_1(\lambda)=\frac{\mathcal{N}_{\rm TN}(\lambda+i\eta/2)\mathcal{N}_{\rm TN}(\lambda-i\eta/2)}{\chi_{\rm TN}(\lambda)}-1\,,
\label{eta_tilted_neel}
\ee
where the functions $\chi_{\rm TN}(\lambda)$ and $\mathcal{N}_{\rm TN}(\lambda)$ are given in  \eqref{chinondiag} and \eqref{normnondiag} respectively. 

After rewriting $\zeta$ in terms of $\vartheta$ as in \eqref{def_zeta}, it is straightforward to see that \eqref{eta_tilted_neel} coincides with the expression for $\eta_1(\lambda)$ recently derived in \cite{pvc-16} for the tilted N\'eel state. Note that a completely different approach was used in \cite{pvc-16}, which was based on the knowledge of all the quasi-local charges of Hamiltonian \eqref{hamiltonian} \cite{iqdb-16}. An analogous calculation, exploiting the identifications \eqref{fpar1}-\eqref{fpar4}, shows that the same derivation also yields the function $\eta_1(\lambda)$ for the tilted ferromagnet, recovering also in this case the result presented in \cite{pvc-16}.
The explicit identification exploited in this work also explains the
validity of the $Y$-system relations between the functions
$\eta_n(\lambda)$: they follow from the fusion hierarchy of the
relevant QTM's. Note that prior to this work,
the $Y$-system for the tilted N\'eel and tilted ferromagnet states was only verified
numerically in \cite{pvc-16, iqdb-16}. 

We can also briefly comment on the relation between the quantum transfer matrix approach to the Loschmidt echo and the generalized Gibbs ensemble (GGE) construction, mentioned in the introduction. For the states considered in this work, the distribution functions $\eta_n(\lambda)$ associated with the GGE coincide with those of the representative eigenstates derived in \cite{pvc-16}, and hence with those obtained as the solution of our integral equations in the limit $w\to 0$. However, this is no longer true for $w\neq 0$, as the solutions of our integral equation depend on $w$, while the distributions of the GGE do not. This is not surprising as the latter ensemble does not necessarily reproduces the exact result for the Loschmidt echo at large times, as shown in \cite{fagotti2-13}.

It is an interesting question whether the overlaps with the Bethe eigenstates can be recovered
from the generalized TBA equations \eqref{TBA_nondiagonal}.
At present such overlaps are only known for the states described by
diagonal K-matrices. However, even in the non-diagonal case it is possible to reconstruct the
extensive part requiring that the quench action method results in the
source terms of \eqref{TBA_nondiagonal}. We plan to return to this
question in future works.

To conclude this section, we stress that the identification of the QA and QTM descriptions only
applies to real $\beta$ parameters. There are two main reasons for
this. First, the quench action itself has to be strictly real,
and this only happens for real $\beta$. 
An analytic continuation to complex $\beta$ is expected to
give correct results in some neighbourhood of the real axis, but it is
not clear whether the QA method could account for the physical
non-analyticities caused by the possible level crossings of the quantum
transfer matrix. The second reason for the limitation to real $\beta$
is the fact that the QTM method has been established for the pure
N\'eel state, whereas the quench action method deals with the symmetric
combination \eqref{eq:symmetric_neel}. For real $\beta$ the two states
lead to the same Loschmidt amplitude, similar to what was observed in
the XX model in \cite{as-14}. However, this is not true for generic
$\beta$. Regarding the symmetric state the QTM method has to be
extended to described the transition amplitude \eqref{eq:transition}
as well. This can be achieved with proper modification of our
construction, but it is out of the scope of the present paper and it
is left for future research.

\section{Conclusion}\label{sec:conclusion}

We have considered the analytical computation of the Loschmidt echo following a quantum quench to the XXZ Heisenberg spin chain. We employed a quantum transfer matrix approach, focusing on generic two-site product initial states. In the case of tilted N\'eel states, we have provided explicit results in terms of the solution of TBA-like integral equations, which were derived by fusion of certain boundary transfer matrices. Our work extends the results of \cite{pozsgay-13}, where quenches from the N\'eel state were studied. On the one hand, the integral formulas derived in this work appear to be convenient for numerical evaluation. In particular, in some regimes we were able to follow the post-quench dynamics of the system for very large times. On the other hand, the derivation of TBA-like integral equations from fusion of boundary transfer matrices employed here seems to be easily generalized to the case of generic two-site product states, as we explicitly showed in the case of tilted N\'eel states.

The integral equations derived in this work are valid for arbitrary imaginary times and real times up to a maximum time $t_0$. In some regimes the latter is seen to be either very large or infinite. For times larger than $t_0$ the integral equations have to be modified in a way which is in principle feasible and discussed in the previous sections. A detailed analysis of the modified equations in these cases remains an interesting issue to investigate. For example, it might lead to a fully analytical computation of the points of non-analyticity in the time evolution for quenches within the same quantum phase, as those observed and discussed in \cite{fagotti2-13, vd-14, as-14, ssd-15}.

As an important part of our work, we have also established a clear
correspondence between the quantum transfer matrix formalism and the
quench action approach. This generalizes to non-equilibrium settings
the well-known equivalence of the quantum transfer matrix and
thermodynamic Bethe ansatz methods \cite{tsk-01}. By means
of this correspondence we provided an alternative derivation of the
Bethe ansatz rapidity distribution functions for tilted N\'eel and
tilted ferromagnet states recently obtained in \cite{pvc-16}. In
particular, within our treatment the $Y$-system relations satisfied by
the functions $\eta_n(\lambda)$ of a given state are naturally derived
from the fusion relations of the corresponding boundary transfer
matrices. 

In this work we have focused on the gapped regime of the XXZ chain. A natural question regards the possibility of extending our calculations to the gapless phase. While the quantum transfer matrix formalism employed in this work still applies, we expect that different technicalities will arise in the analytical computation of the leading eigenvalue of the boundary quantum transfer matrix. In particular, restricting to values of the anisotropy corresponding to ``roots of unity" \cite{takahashi-99}, experience with the thermal case  suggests that the final result for the Loschmidt echo will be written in terms of a finite, rather than infinite, set of TBA-like integral equations. This represents an intriguing direction to explore, especially in connection to recent studies regarding the physical implication of additional conservation laws in the gapless regime for quenches from initial states breaking spin-inversion symmetry \cite{prosen-11,prosen-14,impz-16, dcd-16}.

It is also interesting to wonder whether the method employed here can be
generalized to compute the Loschmidt echo from $k$-site product
initial states with $k>2$. By following our derivation, it is clear
that these should be associated to boundary transfer matrices with
$k>2$ auxilliary rows. While such transfer matrices could in fact be
obtained by fusion of the elementary 
 boundary transfer matrices considered in the present
 work\footnote{This idea was suggested to us by Frank G\"ohmann.}, a
 dedicated study is needed to establish whether 
 these might actually correspond to some non-trivial physical states.
On the other hand, the observed failure of the $Y$-system for certain
 initial states  \cite{pvc-16,iqdb-16,iqc-16} might be a sign that there is no family
 of commuting QTM's describing such states. We leave the
 investigation of these questions to a future work.

Finally, a natural question is whether techniques similar to those
employed in this work might be used to provide analytical formulas for
the time evolution of local observables. While the quantum transfer
matrix formalism seems to give a convenient framework to tackle this
problem, the latter remains very challenging
as already noted in
\cite{pozsgay-13}. Nevertheless, we hope that our results might
motivate further investigations in this direction .

\section{Acknowledgments}
We are very grateful to Pasquale Calabrese for collaboration at an
early stage of this project and many discussions. We thank G\'{a}bor
Tak\'{a}cs for providing numerical data to preliminary test the
validity of our numerical evaluation of the Loschmidt echo, and for
very useful comments. In particular, we thank him for pointing out
that additional poles arise also in the solution of the TBA equations
for the case of the N\'eel state for sufficiently small values of the
anisotropy. B. P. acknowledges interesting discussions with Frank
G\"{o}hmann, Andreas Kl\"{u}mper and Junji Suzuki. Also, B. P. is
grateful to Pasquale Calabrese and SISSA for their hospitality.
E. V. acknowledges
support by the ERC under Starting Grant 279391
EDEQS. B. P. acknowledges support from the ``Premium'' Postdoctoral
Program of the Hungarian Academy of Sciences, and the NKFIH grant no. K 119204.

\appendix

\section{Analytic solution of the Bethe equations for $2N=2$}\label{app:bethe_eq}

In this appendix we analytically determine the solution of the Bethe equations \eqref{eq:inh_bethe_eq} associated with the leading eigenvalue of the transfer matrix \eqref{QTMdef} for $2N=2$. We focus on the non-diagonal case associated to the tilted N\'eel state in the limit $\beta\to 0$. This allows to analytically verify the relation \eqref{lambda0} in the case $2N=2$ when $\beta\to 0$.

For Trotter number $2N=2$, the Bethe roots associated with the leading eigenvalue consist of two purely imaginary regular Bethe roots $\lambda^{\rm reg}_{1,2}=\pm \lambda^{\rm reg}$ and two extra Bethe roots  $\lambda^{\rm extra}_{1,2}=\pm \lambda^{\rm extra}$, cf. section~\ref{sec:tNeelQTM}. In the limit $\beta\to 0$ the regular roots approach zero, namely $\lambda^{\rm reg}\to 0$, while $\lambda^{\rm extra}\to x$ where $x$ is a real number different from $0$. The set of doubled Bethe roots in the limit $\beta\to 0$ then reads
\be
\lim_{\beta\to 0}\{\tilde{\lambda}_{j}\}_{j=1}^4=\{0,0,-x,x\}\,.
\label{doubled_set}
\ee
We now determine the value of $x$ analytically. Substituting \eqref{doubled_set} for the doubled rapidities in the Bethe equations \eqref{eq:inh_bethe_eq} and choosing the boundary parameters as in \eqref{par_1}-\eqref{def_zeta}, we obtain the following equation for $x$
\bea
\fl -\sinh (\eta ) \sinh (x-\eta ) \cosh \left(-\zeta -\frac{\eta }{2}+x\right) \cosh \left(\zeta -\frac{\eta }{2}+x\right)\nonumber\\
+\sinh (\eta ) \sinh (\eta +x) \cosh \left(-\zeta +\frac{\eta }{2}+x\right) \cosh \left(\zeta
   +\frac{\eta }{2}  +x\right)\nonumber\\
   +\frac{1}{2} [1-\cosh (3 \eta )] \sinh (x) \sinh (2 x)=0\,.
   \label{aux_intermediate_equation}
\eea
Here we already excluded the cases $x=0$, $x=\eta/2$, $x=\eta$ which do not correspond to the leading eigenvalue. One can now express the l. h. s. of \eqref{aux_intermediate_equation} as a trigonometric polynomial; substituting then
\be
x\to\frac{1}{2}\log z\,,
\label{correspondence}
\ee
the l. h. s. is transformed into a polynomial in powers of $z$. The corresponding roots $z_j$ can be easily found analytically, which in turn give the corresponding solutions $x_j$ of \eqref{aux_intermediate_equation} through \eqref{correspondence}. It is found that there is only one real positive solution of \eqref{aux_intermediate_equation}, which is given by
\bea
\fl x= \frac{1}{2} \log \left\{4 \cosh ^2(\zeta ) (\cosh (\eta )+1)
 +\cosh (2 \eta )+
  2 \sqrt{2} \cosh\left(\frac{\eta }{2}\right) \right.  \nonumber\\
\fl \times \left. \sqrt{ [\cosh (2 \zeta )
+\cosh (\eta )]\left[2 \cosh ^2(\zeta )
   (\cosh (\eta )+1)+\cosh ^2(\eta )\right]} \right\} \,.
\label{final_x}
\eea

One can now directly substitute \eqref{doubled_set} [with the explicit value of $x$ given in \eqref{final_x}] into \eqref{TQrelation_nondiagonal} which, using \eqref{BQTMlambdadef}, directly yields the final expression for $\Lambda_0(u)$. The calculations are tedious but are easily performed using the program Mathematica. The final resut reads
\be
\Lambda_0(u)\equiv 1\,,
\ee
as in \eqref{lambda0}.

\section{Compact expression for the dynamical free energy}\label{app:dyn_free_en}

In this appendix we show the equivalence between the expressions \eqref{eq:temp_g} and \eqref{QA_final_g}. We start by presenting a number of identities which will be needed in the derivation. First, for any pair of functions $f(\lambda)$, $g(\lambda)$ one has
\be
\int_{-\pi/2}^{\pi/2}{\rm d}\lambda f(\lambda)g(\lambda)=\frac{1}{\pi}\sum_{k\in\mathbb{Z}}\hat{f}(k)\hat{g}(-k)\,,
\label{identity_1}
\ee
where we used the conventions \eqref{fourier_1}, \eqref{fourier_2} for the Fourier transform. Next, the partially decoupled form of the Bethe equations \eqref{eq:thermo_bethe} yields the following identity for the Fourier transform of $\rho_n(\lambda)$ and $\rho^h_n(\lambda)$ \cite{takahashi-99}
\be
\hat{\rho}_n(\kappa)+\hat{\rho}^h_n(\kappa)=\frac{1}{2\cosh(k\eta)}\left[\hat{\rho}^h_{n-1}(k)+\hat{\rho}^h_{n+1}(k)\right]\,,
\label{identity_2}
\ee
where we used the convention
\be
\hat{\rho}^h_0(k)=1\,.
\ee
Analogously, from the decoupled form \eqref{eta_TBA} one obtains
\be
\reallywidehat{\log \eta_1}(k)=-\frac{2\pi\beta}{\cosh(k\eta)}+\hat{d}_1(k)+\frac{1}{2\cosh(k\eta)}\reallywidehat{\log\left(1+\eta_{2}\right)}\,.
\ee
and
\be
\fl \reallywidehat{\log \eta_j}(k)
=\hat{d}_n(k)+\frac{1}{2\cosh(k\eta)} \left[
\reallywidehat{\log\left(1+\eta_{j+1}\right)(k)} 
+
\reallywidehat{\log\left(1+\eta_{j-1}\right)}(k)  
 \right]\,,\quad n\geq 2\,,
\label{identity_3}
\ee
where we introduced the parameter $\beta=\beta_{w}$ defined in \eqref{beta_parameter}.

Next, by making explicit use of the Fourier transform for the function $a_n(\lambda)$ defined in Eq.~\eqref{def:a_function}
\be
\hat{a}_n(k)=e^{-|k|n\eta}\,,
\ee
one can verify the identities \cite{takahashi-99}
\be
\fl (a_0+a_2)\ast a_{nm} = a_1\ast (a_{n-1,m}+a_{n+1,m}) + (\delta_{n-1,m}+\delta_{n+1,m})\,a_1\,, \quad n\geq 2,\, m \geq 1 \,,
\label{a_identity_1}
\ee
and
\be
(a_0+a_2)\ast a_{1,m} = a_1 \ast a_{2,m} + a_1 \, \delta_{2,m}\,, \quad m\geq 1 \,,
\label{a_identity_2}
\ee
where we defined $a_0(\lambda) = \delta(\lambda)$. Using now \eqref{a_identity_1}, \eqref{a_identity_2} and convolving \eqref{eq:saddle-point} with $(a_0+a_2)$ one can derive the identities
\bea
 (a_0&+&a_2)\ast\log\eta_1 =-4\pi\beta a_1 -(a_0+a_2)\ast \log W_1 + a_1\ast \log W_{2} \nonumber\\
 &+& a_1 \ast \log(1+\eta_{2})\,,\label{aux_eq_4a}\\
 (a_0&+&a_2)\ast\log\eta_n = -(a_0+a_2)\ast \log W_n + a_1\ast(\log W_{n-1}+\log W_{n+1}) \nonumber\\
 &+& a_1 \ast \big[\log(1+\eta_{n-1})+\log(1+\eta_{n+1})\big]\,,\qquad n\geq 2\,.
\label{aux_eq_4b}
\eea
Taking the Fourier transform of \eqref{aux_eq_4a}, \eqref{aux_eq_4b} one obtains directly
\bea
&-&\reallywidehat{\log W_1}+\frac{1}{2\cosh(k\eta)}\reallywidehat{\log W_{2}}=\nonumber\\
&=&\frac{4\pi\beta}{2\cosh(k\eta)} +\reallywidehat{\log\eta_1}-\frac{1}{2\cosh(k\eta)}
\reallywidehat{\log\left(1+\eta_{2}\right)} \,,\label{new_aux_1}
\eea
\bea
&-&\reallywidehat{\log W_n}+\frac{1}{2\cosh(k\eta)}\left[\reallywidehat{\log W_{n-1}}+\reallywidehat{\log W_{n+1}}\right]=\nonumber\\
&=&\reallywidehat{\log\eta_n}-\frac{1}{2\cosh(k\eta)} \left[
\reallywidehat{\log\left(1+\eta_{n+1}\right)} 
+
\reallywidehat{\log\left(1+\eta_{n-1}\right)}  
 \right]\,.\label{new_aux_2}
\eea

We are now ready to rewrite \eqref{eq:temp_g}. First, by means of \eqref{identity_1} and making repeated use of \eqref{identity_2} one has

\bea
\fl S_{\rm YY}[\brho^*]=\frac{1}{2\pi}\sum_{k\in \mathbb{Z}}\frac{1}{2\cosh(k\eta)}\reallywidehat{\log\left(1+\eta_1\right)}(-k)\nonumber\\
\fl + \frac{1}{2\pi}\sum_{k\in \mathbb{Z}}\sum_{n=1}^{\infty}\rho_n^{h}(k)\left[\frac{1}{2\cosh(k\eta)} \left(\reallywidehat{\log(1+\eta_{n-1})}(-k) +\reallywidehat{\log(1+\eta_{n+1})}(-k)\right)\right.\nonumber\\
-\left. \reallywidehat{\log(\eta_{n})}(-k)\right]\,.
\label{aux_eq_5}
\eea
Next, employing \eqref{identity_1} and making again repeated use of \eqref{identity_2} the functional $S_{\rm Q}$ is recast in the form
\bea
\fl 2S_{\rm Q}[\brho^*]=-\frac{1}{2\pi}\sum_{k\in\mathbb{Z}}\frac{1}{2\cosh(k\eta)}\reallywidehat{\log W_1}(-k)+\nonumber\\
\fl \frac{1}{2\pi}\sum_{k\in\mathbb{Z}}\sum_{n=1}^{\infty}\rho^h_{n}(k)\left[\reallywidehat{\log W_n}(-k)-\frac{1}{2\cosh(k\eta)}\left(\reallywidehat{\log W_{n-1}}(-k)+\reallywidehat{\log W_{n+1}}(-k)\right)\right]\,,
\eea
which can be directly rewritten using \eqref{new_aux_1}, \eqref{new_aux_2} as
\bea
\fl 2S_{\rm Q}[\brho^*]=-\frac{1}{2\pi}\sum_{k\in\mathbb{Z}}\frac{1}{2\cosh(k\eta)}\reallywidehat{\log W_1}(-k)-4\pi\beta\frac{1}{2\pi}\sum_{k\in\mathbb{Z}}\rho^h_1(k)\frac{1}{2\cosh(k\eta)}\nonumber\\
\fl + \frac{1}{2\pi}\sum_{k\in \mathbb{Z}}\sum_{n=1}^{\infty}\rho_n^{h}(k)\left[\frac{1}{2\cosh(k\eta)} \left(\reallywidehat{\log(1+\eta_{n-1})}(-k) +\reallywidehat{\log(1+\eta_{n+1})}(-k)\right)\right.\nonumber\\
-\left. \reallywidehat{\log(\eta_{n})}(-k)\right]\,.
\eea

Finally, in the same way the energy $e[\brho^*]$ can be rewritten as
\be
\fl we[\brho^\ast]=-\pi Jw\sinh(\eta) \frac{1}{\pi}\sum_{k\in\mathbb{Z}}\frac{e^{-|k|\eta}}{2\cosh(k\eta)}+2\pi\beta \frac{1}{\pi}\sum_{k\in\mathbb{Z}}\rho_1^{h}(k)\frac{1}{2\cosh(k\eta)}\,,
\ee
where we used again \eqref{beta_parameter}. Putting everything together, Eq.~\eqref{eq:temp_g} is rewritten as
\bea
g(w)=\frac{1}{2\pi}\sum_{k\in \mathbb{Z}}\frac{1}{2\cosh(k\eta)}\reallywidehat{\log\left(1+\eta_1\right)}(-k)\nonumber\\
+\frac{1}{2\pi}\sum_{k\in\mathbb{Z}}\frac{1}{2\cosh(k\eta)}\reallywidehat{\log W_1}(-k)+\pi Jw\sinh(\eta) \frac{1}{\pi}\sum_{k\in\mathbb{Z}}\frac{e^{-|k|\eta}}{2\cosh(k\eta)}\,,
\eea
namely
\be
\fl g(w)=\frac{1}{2}\int_{-\pi/2}^{+\pi/2}\,d\mu\ s(\mu)\left\{2\pi w J\sinh(\eta) a_1(\mu)+ \log\left[1+\eta_1(\mu)\right]+\log W_1(\mu)\right\}\,.
\label{final_formula}
\ee
It remains to show
\be
\int_{-\pi/2}^{+\pi/2}\,d\mu\ s(\mu)\log W_1(\mu)=-\int_{-\pi/2}^{+\pi/2}\,d\mu\ s(\mu)\log \left[1+Y_1(\mu)\right]\,,
\label{to_prove}
\ee
where $Y_1(\mu)$ is given in \eqref{definition_y1}, with the definitions \eqref{chidiag} and \eqref{normdiag} valid for the N\'eel state. In order to do so, we first rewrite
\bea
1+Y_1(\lambda)&=& \frac{4\sin ( 2 \lambda +i\eta) \sin ( 2 \lambda -i\eta)}{\sin^2\left( \lambda +i\frac{\eta }{2}\right)\sin^2\left( \lambda -i\frac{\eta }{2}\right)}\nonumber\\
&\times &\frac{\sin \left[\lambda+i (\Xi+\eta )/2 \right] \sin \left[\lambda-i (\Xi+\eta )/2 \right]}{\sin(2 \lambda +2i \eta )\sin(2 \lambda -2i \eta )} \nonumber\\
&\times & \sin \left[\lambda+i (\Xi-\eta )/2 \right] \sin \left[\lambda-i (\Xi-\eta )/2 \right]\,,
\eea
where we introduced the parameter $\Xi$ which is defined through
\be
\cosh\Xi=\cosh^2\eta\,.
\ee
The equality \eqref{to_prove} can then be established directly by explicit calculation using that for a function $f(\lambda)$ with no poles in the physical strip \eqref{physicalstrip} the following identity holds
\be
\int_{-\pi/2}^{+\pi/2}\,d\mu\ s(\lambda-\mu)\left[f(\mu-i\eta/2)+f(\mu+i\eta/2)\right]=f(\lambda)\,.
\ee
Equations \eqref{final_formula} and \eqref{to_prove} finally yield \eqref{QA_final_g}.

\end{document}